\documentclass[twocolumn,aps,prd,amsmath,amssymb,preprintnumbers,nofootinbib]{revtex4-1}
\usepackage{graphicx,caption}
\usepackage{subcaption}
\usepackage{soul}
\usepackage{hyperref,url}
\hypersetup{
	colorlinks=true,       
	linkcolor=blue,          
	citecolor=blue,        
}

\begin{document}
\setcounter{footnote}{0}

\title{Scalar quasinormal modes in 
emergent modified gravity}
\author{Martin Bojowald}
\email{bojowald@psu.edu}
\affiliation{Institute for Gravitation and the Cosmos,
The Pennsylvania State University, 104 Davey Lab, University Park, PA 16802, USA}

\author{Erick I.\ Duque}
\email{eqd5272@psu.edu}
\affiliation{Institute for Gravitation and the Cosmos,
The Pennsylvania State University, 104 Davey Lab, University Park, PA 16802, USA}

\author{S. Shankaranarayanan}
\email{shanki@iitb.ac.in}
\affiliation{Department of Physics,  Indian Institute of Technology Bombay, Mumbai 400076, India}

\begin{abstract}
Emergent modified gravity is a post-Einsteinian gravitational theory where spacetime geometry is not fundamental but rather emerges from the gravitational degrees of freedom in a non-trivial way. The specific relationship between geometry and these degrees of freedom is unique for each theory, but it is not predetermined. Instead, it is derived from constraints and equations of motion, relying on key aspects of the canonical formulation of gravity, such as structure functions in Poisson brackets of constraints and covariance conditions.
As shown in previous work, these new theories allow for two types of scalar matter coupling: (1) minimal coupling, where the matter equations of motion mirror the Klein-Gordon equation on a curved emergent spacetime, and (2) nonminimal coupling, where the equations deviate from the Klein-Gordon form but still respect covariance.
Observable features, such as the quasinormal mode spectrum, can help distinguish between different couplings based on how well their predictions match the data. In this work, the spectra of scalar quasinormal modes for both minimal and nonminimal couplings are derived using the third-order WKB approximation. Significant differences are found between the two cases. Notably, the nonminimal coupling allows for vanishing real and imaginary frequency components, and even opposite-sign values for the imaginary part at sufficiently small mass scales, pointing to potential new physical implications.
Finally, the high-frequency QNM spectra in emergent modified gravity is identical to the classical result, up to an overall constant, suggesting that the horizon area spectrum remains equispaced.
\end{abstract}

\maketitle

\section{Introduction}

Several predictions of general relativity (GR) have enjoyed exceptional success in modern astronomical observation tests, including the detection of gravitational waves (GW) among the most recent and remarkable ones~\cite{GW1,GW2,GW_Review}.
GWs emitted during the ring-down phase are known as quasinormal modes (QNMs) \cite{ReggeWheeler,Zerilli1,Zerilli2,ChandraQNM} and provide crucial insights into the nature of the emitting objects~\cite{GW_Review,Cardoso_2019,Baibhav_2023}.
By extracting the QNM frequency and damping time, one can conduct various tests of gravity~\cite{Cardoso_2019,Baibhav_2023,GW170104,TestofGR,ToG,GWTemplates} and examine the no-hair theorem~\cite{NoHair}. Such analyses have been performed on the remnant black holes formed from compact mergers detected by the Advanced LIGO and Virgo collaborations~\cite{TestofGR,ToG}.
The amplitude of the GW ringdown signal scales with the mass of the remnant black hole, while its frequency decreases.
Therefore, these tests are particularly sensitive when the remnant black hole is large and with more advanced detectors with lower frequency bands.
The upcoming third-generation gravitational-wave detectors, such as the ground-based Cosmic Explorer ($\sim$1Hz--5kHz) \cite{Evans:2016mbw,Evans:2023euw} and the Einstein Telescope ($\sim$1Hz--10kHz) \cite{Maggiore:2019uih,Branchesi:2023mws}, as well as the space-based LISA ($\sim$0.1mHz--1Hz) \cite{LISA,Barausse:2020rsu,LISA:2022kgy}, will offer significantly higher signal-to-noise ratios (SNR) in the quasi-normal mode regime, with ${\rm SNR} > 50$~\cite{QNMSNRbound}, allowing for more precise probing of QNM structures.
The frequency bands of the Cosmic Explorer and the Einstein Telescope are better suited for stellar-mass binaries, while LISA will have the capability to analyze supermassive or intermediate-mass black holes.

As the accuracy of these observations increases, the possibility of testing GR in strong or weak gravity regimes grows ever more likely, promising a set of restrictions on potential deviations from GR. Possible modified gravity theories will thus be subjected to these constraints, resulting in an objective procedure to falsify them. These expectations require an excellent class of theories beyond GR, subjected to the stringent conditions of general covariance and stability and general observational statements such as the close proximity of the speeds of gravitational and electromagnetic waves. 

The need for alternative theories of gravity is motivated by theory and observations~\cite{Shanki-Joseph:Review}. Applications of GR to cosmology and black holes in the low and strong curvature regimes have resulted in several unexplained phenomena, including the Hubble tension, the cause of late-time acceleration, and the dark matter problem.
Furthermore, GR robustly predicts the formation of singularities at the centers of black holes and the Big Bang. In the standard Lagrangian formulation, modified gravity is generally understood as the addition of higher-curvature terms to the Lagrangian density, a scalar resulting from different contractions of the Riemann tensor; hence, covariance is built in. However, such higher-curvature terms typically result in instabilities and introduce extra degrees of freedom~\cite{Shanki-Joseph:Review}.
The QNM spectrum predicted by GR consists of two isospectral towers of modes, which are even and odd under parity~\cite{GW_Review}. In contrast, alternative theories of gravity can introduce additional polarization modes and affect QNMs in three distinct ways~\cite{2018-Barack.etal-CQG}: 
\begin{enumerate}
\item Alter the spectrum of both even and odd modes while maintaining isospectrality. 
\item Break isospectrality, resulting in unequal gravitational wave emission between the two polarization states~\cite{Shanki_QNM_fR,Shanki-QNM-dCS,Shanki-Essay:2019}. 
\item Mix even and odd modes, eliminating the distinction between them.
\end{enumerate}
This prompts the question of whether there are other types of modified gravity beyond simply adding higher-curvature terms or beyond any class of covariant modified theories based on space-time scalars that circumvent the aforementioned challenges. Emergent modified gravity (EMG) is one such alternative \cite{EMG}.

EMG is a canonical formulation of gravity and incorporates the idea that the spacetime geometry is not a presupposed object but rather derived from a set of covariance conditions \cite{EMGcov}. An explicit distinction is made between the fundamental gravitational field (defined as the set of degrees of freedom composing the phase space) and the spacetime metric field, which provides the geometrical content. The latter is treated as an emergent object that depends on the gravitational field but is not necessarily identical to it. Applications of EMG in spherical symmetry include nonsingular black hole solutions in vacuum \cite{alonso2022nonsingular,Alonso_Bardaji_2022,ELBH}, the formation of wormholes or black-to-white-hole transitions as a result of gravitational collapse \cite{EMGPF}, and a relativistic realization of modified Newtonian dynamics (MOND) \cite{milgrom1983modification,banik2022galactic,MONDEMG}.

An example of a new dimensionless modification function allowed by EMG is of the form $\lambda(q_{\vartheta\vartheta})$, a general and arbitrary function of the angular part  $q_{\vartheta\vartheta}$ of a spherically symmetric spatial metric $q_{ab}$. 
This function is responsible for the nonsingular behavior of black-hole solutions, the existence of a minimum radius, and MOND-like effects when it is logarithmic.
These modifications do not depend on curvature invariants but rather on the squared areal radius and, therefore, may vary with the radius around a black hole.
The general formulation of EMG does not offer a specific choice for the dependence $\lambda(q_{\vartheta\vartheta})$ on theoretical grounds, just as, say, the general set of $f(R)$ theories does not fix the function $f$. Hence, this function may be chosen to accommodate specific phenomenological properties. (In addition, detailed studies of canonical quantum gravity that lead to EMG-type effective descriptions may impose independent conditions on the modification functions.)

Since GR is recovered in the limit $\lambda\to0$, it is often suggested that a decreasing function should be chosen, for instance, $\lambda=\sqrt{\Delta/q_{\vartheta\vartheta}}$ where $\Delta$ is a constant with units of area, such that it is asymptotically vanishing and the classical space-time is recovered at large distances.
Another popular choice is a constant $\lambda$, mainly because it simplifies calculations.
However, there are two drawbacks related to the choice of constant $\lambda$: 1) The function $\lambda$ is unitless and hence does not scale with the system at hand, neither on distance scales nor on mass scales, and 2) in the presence of a cosmological constant, a constant $\lambda$ predicts a maximum radius of the universe \cite{alonsobardaji2023Charged}. The given choice of a decreasing $\lambda(q_{\vartheta\vartheta})$ is free of both problems as it contains the constant $\Delta$ with units of area and asymptotically recovers the (anti-)de Sitter geometry \cite{ELBH}. The function, therefore, is not arbitrary, but freedom remains, for instance, in the precise fall-off behavior for large radii or in its small-radius form if it is not a complete power law. A more reliable method to restrict possible choices of this function and its parameters is through observations. 
The best fit to multiple observational tests may then indicate a more complete form for this function.

In addition to implying the existence of a new modification function $\lambda$, EMG admits different versions of covariant couplings for matter fields with possible modifications of their own \cite{EMGPF,alonso2021anomaly,EMGscalar,EmEM}.
In particular, scalar matter can be coupled in different inequivalent ways \cite{EMGscalar}.
One class of scalar matter coupling, referred to as minimal coupling \cite{MinCoup}, consists in simply using the expression of the emergent metric in the Klein--Gordon equation.
The second class of coupling, which we will correspondingly refer to as nonminimal coupling, results in more complicated matter equations of motion that do not resemble the Klein-Gordon equation. But characteristic classical symmetries, for instance covariance and the conserved current of a free field, are still respected.
On theoretical grounds, the latter kind of modification is preferred because it implies a nonsingular homogeneous solution, while the former always predicts the formation of a singularity in the presence of matter \cite{EMGscalar}.
Here we are interested in further discrepancies in the physical predictions of minimal vs.\ nonminimal couplings, and of constant vs.\ decreasing $\lambda$ functions in the context of scalar QNMs.

Scalar modes are often considered less relevant for current astrophysical observations because they do not directly describe the tensor modes of spacetime perturbations that produce observable gravitational waves. However,
gravitational waves are considerably more complicated to study in EMG and their full treatment has not been formulated yet. This is because of the tensorial aspect of the gravitational perturbations that make the calculations for the closure of brackets very lengthy. Hence, as a first step, we do the analysis for the scalar QNMs, with the idea that all the 
important features of scalar QNMs will carry forward to gravitational waves.
The minimally coupled case with constant $\lambda$ was already studied in \cite{Fu_2023,Moreira_2023,Gingrich_2024}.
Here, we extend these results to a decreasing parameter of the form $\lambda=\sqrt{\Delta/q_{\vartheta\vartheta}}$ with a constant $\Delta$ with dimensions of area that allows for scale dependent results that are impossible for the case of constant and dimensionless $\lambda$.
We also compare these results with nonminimally coupled models.
The resulting spectrum differs significantly in each case, which strongly suggests that the observation of QNM spectra can be a good candidate to test and falsify models of EMG. Conversely, EMG appears to offer more freedom to model different QNM spectra compared with traditional alternative theories of gravity.
As a qualitative statement, this result from scalar quasinormal modes is likely to hold also for their more complicated tensorial counterpart.

In Section~\ref{sec:Classical scalar QNMs}, we briefly review the canonical derivation for QNMs of scalar matter.
In Section~\ref{sec:Spherical emergent modified gravity} we review EMG in spherical symmetry, the vacuum background solution, the (non)minimal coupling of scalar matter, as well as a proof that there is no backreaction for test fields, and derive the relevant equations of motion for the QNMs.
In Section~\ref{sec:Quasinormal modes spectra} we use the WKB approximation to third order to solve the QNM spectra of minimally and nonminally coupled scalar matter for both constant and decreasing $\lambda$, and we present the spectra for the $n=0,1,2$ modes with $l=0,1,2$.
Since multiple versions of EMG are available and there are no current restrictions from observations, we attempt to give a broad classification of possible outcomes for various choices of modifications.
Finally, Section~\ref{sec:Discussion} presents a discussion of the results and possible implications for low mass black holes.
Appendices A and B contain the details of the calculations presented in the main text.

We use $(-,+,+,+)$ signature for the 4-D space-time metric. Lower-case Greek (Latin) alphabets denote the 4-D space-time (3-space) coordinates. We set $G = c = 1$. A dot denotes derivative with respect to the time coordinate ($t$), while a prime denotes derivative with respect to the radial coordinate ($x$).

\section{Quasinormal modes of classical scalar matter}
\label{sec:Classical scalar QNMs}

We begin with a brief review of the classical treatment of scalar quasinormal modes and then describe its canonical derivation as well as new features in emergent modified gravity.

\subsection{Klein--Gordon equation}

The Klein--Gordon equation on a curved spacetime with metric $g_{\mu\nu}$ is given by
\begin{eqnarray}
    g^{\mu\nu} \nabla_\mu \nabla_\nu \phi 
    - \frac{\partial V}{\partial\phi} = 0
\end{eqnarray}
with the scalar potential $V(\phi)$. On the Schwarzschild background 
\begin{eqnarray}
\label{eq:Schwarzschild}
{\rm d} s^2 &=& - f(x) {\rm d} t^2 + \frac{1}{f(x)} {\rm d} x^2 + x^2 {\rm d} \Omega^2\,;\\
f(x) &=& 1-\frac{2M}{x}\,,\nonumber
\end{eqnarray}
the above equation takes the form
\begin{equation}\label{eq:KG on Schwarzschild}
    - \frac{\ddot{\phi}}{f(x)}
+ \frac{2}{x}\left(1-\frac{M}{x}\right)\  \partial_x\phi
    + f(x) \partial_x^2\phi
    + \Delta^{\vartheta} \phi
    - \frac{\partial V}{\partial\phi}
    = 0\,,
\end{equation}
where
\begin{eqnarray}
    \Delta^\vartheta = \frac{1}{x^2\sin \vartheta}\partial_\vartheta\left[\sin \vartheta \partial_\vartheta \right]
    + \frac{\partial_\varphi^2}{x^2\sin^2\vartheta}
\end{eqnarray}
is the Laplacian on the 2-sphere.

In a spherically symmetric background, it is useful to perform the expansion
\begin{equation}\label{eq:Spherical harmonics exp}
    \phi (t,x,\vartheta,\varphi) = \sum_{l,m} \tilde{\phi}_{lm}(t,x) Y_{l m}(\vartheta,\varphi)
\end{equation}
in terms of the \emph{real} spherical harmonics $Y_{l m}(\vartheta,\varphi)$.
%
Using the property
\begin{eqnarray}
    \Delta^\vartheta Y_{l m}(\vartheta,\varphi) = - \frac{l(l+1)}{x^2}
\end{eqnarray}
and a vanishing potential $V=0$ for a free field, the Klein--Gordon equation simplifies to

\begin{eqnarray}\label{eq:KG on Schwarzschild-Veff}
    - \frac{1}{f(x)} \ddot{\tilde{\phi}}_{lm}
    + \frac{2}{x}\left(1-\frac{M}{x}\right)\ \partial_x \tilde{\phi}_{lm}
    \qquad&&
    \nonumber\\
    + f(x) \partial_x^2 \tilde{\phi}_{lm}
    - l(l+1) \tilde{\phi}_{lm}
    &=& 0
\end{eqnarray}
for each mode.
We find that this equation is precisely the spherically symmetric Klein-Gordon equation (\ref{eq:KG on Schwarzschild}) with an effective potential
\begin{eqnarray}\label{eq:Effective potential angular momentum}
    V_{\rm eff} (\tilde{\phi}_{l}) = \frac{l(l+1)}{2x^2}\tilde{\phi}_{lm}^2\,,
\end{eqnarray}
with $\tilde{\phi}_{lm}$ replacing $\phi$. Alternative theories of gravity such as EMG may have more complicated equations of motion compared to the standard Klein--Gordon equation.
However, as long the equations of motion are linear in $\phi$, the spherical harmonics expansion is viable and we could work entirely in spherical symmetry by just applying this potential rather than setting $V=0$ for simplicity, avoiding the more complicated nonspherical equations. This is the approach we will take in the following sections.

The mode equation can be further simplified by defining
\begin{equation}
    \tilde{\phi}_{lm} (t,x) = \frac{u_{lm} (t,x)}{x}\,,
    \end{equation}
Fourier-transforming in time to $\tilde{u}_l(\omega,x)$ in
\begin{equation}\label{eq:Fourier transform - t}
    u_{lm} (t,x) =\int^\infty_{-\infty} \frac{{\rm d} \omega}{2\pi} \tilde{u}_{lm} (\omega,x) e^{-i\omega t}\,,
\end{equation}
and transforming to a new (so-called tortoise) radial coordinate 
\begin{equation}
    x^*= \int \frac{{\rm d}x}{f(x)} = 
    x+2M\ln\left(\frac{x}{2M}-1\right)\,,
\end{equation}
where we used the metric components of the Schwarzschild background.
The mode equation then becomes
\begin{equation}
    \frac{{\rm d}^2 \tilde{u}_{lm}}{({\rm d} x^*)^2}
    + \left[\omega^2-V_l(x)\right] \tilde{u}_{lm}
\end{equation}
with
\begin{equation}\label{eq:Classical potential}
    V_l (x)= f(x) \left(\frac{l(l+1)}{x^2}+\frac{2M}{x^3}\right)\,.
\end{equation}

To go about studying the effects of EMG on the QNMs, we need to develop the canonical formalism. In the following subsection, we introduce the  general (non-perturbative) set-up of obtaining QNMs in the canonical formalism of a test scalar field in the Schwarzschild spacetime~\eqref{eq:Schwarzschild}.

\subsection{Quasinormal modes in the canonical formalism}

The action contribution of the scalar matter with potential $V(\phi)$ in a spherically symmetric background has a canonical decomposition given by
\begin{eqnarray}\label{eq:Classical action for scalar field}
    S_{\rm scalar} [\varphi]
    \!\!&=&\!\!
    - \int {\rm d}^4 x\ \sqrt{- \det g} \left[ \frac{\bar{g}^{\mu\nu} (\partial_\mu \phi) (\partial_\nu \phi)}{2}
    + V(\phi)\right]
    \nonumber\\
    \!\!&=&\!\!
    \int {\rm d}^4 x\ \Bigg[
    P_\phi \dot{\phi}
    - N^x H^{\rm scalar}_x
    - N H_{\rm scalar}
    \Bigg]
\end{eqnarray}
with the momentum
\begin{equation}
    P_\phi = \frac{\delta S_{\rm scalar} [\phi]}{\delta \dot{\phi}}
    = \frac{\sqrt{\det \bar{q}}}{N}\; \left(\dot{\phi} - N^a \partial_a \phi\right)\,.
\end{equation}
Here, the spherically symmetric background metric is written in ADM form
\begin{equation}
    {\rm d} s^2 = - \bar{N}^2 {\rm d} t^2 + \bar{q}_{x x} ( {\rm d} x + \bar{N}^x {\rm d} t )^2 + \bar{q}_{\vartheta \vartheta} {\rm d} \Omega^2
    \label{eq:ADM line element - spherical}
    \,,
\end{equation}
with the lapse function $\bar{N}$, the shift vector $\bar{N}^x$, and the spatial metric $\bar{q}_{ab}$.
The barred functions denote background fields independent of the angular coordinates.

The scalar action contains a term $\dot{\phi}P_{\phi}$, which tells us that $\phi$ and $P_{\phi}$ are canonically conjugate with Poisson bracket
\begin{equation}\label{eq:Basic Poisson phi}
    \{\phi(x),P_{\phi}(y)\}=\delta(x-y)\,,
\end{equation}
as well as two contributions proportional to $\bar{N}$ and $\bar{N}^x$, respectively. The latter two are related to energy and momentum densities of the scalar field and can be used to write field equations in canonical form--- 
identified as the Hamiltonian and diffeomorphism constraint contributions of the scalar matter field,
\begin{widetext}
\begin{eqnarray}
    H_{\rm scalar} [\bar{N}]
    &=& \int {\rm d}^3x \bar{N} \Bigg[ \frac{\sqrt{\bar{q}^{xx}}}{2\bar{q}_{\vartheta \vartheta}}\frac{P_\phi^2}{\sin \vartheta}
    + \sin \vartheta\; \bar{q}_{\vartheta \vartheta} \sqrt{\bar{q}^{xx}}\; \frac{(\partial_x\phi)^2}{2}
    + \sin \vartheta\; \bar{q}_{\vartheta \vartheta} \sqrt{\bar{q}_{xx}}\; V(\phi)
    - \frac{\sqrt{\bar{q}_{xx}}}{2} \sin\vartheta\; \phi \tilde{\Delta}^\vartheta \phi
    \Bigg]
    \,,
\end{eqnarray}
\end{widetext}
\begin{equation}
    H_x^{\rm scalar}[\bar{N}^x] = \int{\rm d}^3x \bar{N}^x P_\phi \partial_x\phi
    \,,
\end{equation}
where we introduced
\begin{eqnarray}
    \tilde{\Delta}^\vartheta = \frac{1}{\sin \vartheta}\partial_\vartheta\left[\sin \vartheta \partial_\vartheta \right]
    + \frac{\partial_\varphi^2}{\sin^2\vartheta}\,,
\end{eqnarray}
such that $\Delta^\vartheta=\tilde{\Delta}^\vartheta/x^2$ is the Laplacian on the 2-sphere, and we also defined $\bar{q}^{xx}=1/\bar{q}_{xx}$.
The two first-order equations $\dot{\phi}=\{\phi,H_{\rm scalar}[\bar{N}]+H^{\rm scalar}_x[\bar{N}^x]\}$ and $\dot{P}_{\phi}=\{P_{\phi},H_{\rm scalar}[\bar{N}]+H^{\rm scalar}_x[\bar{N}^x]\}\}$ are equivalent to the second-order Klein--Gordon equation. We will be using these canonical equations because, unlike the action, the constraints require only spatial integrations. They can thus be modified without presupposing the precise form of the spacetime metric, which is instead derived from covariance conditions at a later stage. This is the basis of emergent modified gravity.

Using the basic Poisson bracket (\ref{eq:Basic Poisson phi}), the Hamilton's equations of motion for the scalar field variables in the spherically symmetric background yield
\begin{eqnarray}
    \dot{\phi} &=&
    \frac{\bar{N}\sqrt{\bar{q}^{xx}}}{\bar{q}_{\vartheta \vartheta}}\frac{P_\phi}{\sin \vartheta}
    + \bar{N}^x \partial_x\phi
    \,,\\
    \dot{P}_\phi &=& \sin \vartheta \Bigg[\partial_x \left( \bar{N} \sqrt{\bar{q}^{xx}}\; \bar{q}_{\vartheta \vartheta} (\partial_x\phi)\right)
    + \bar{N} \sqrt{\bar{q}_{xx}}\; \tilde{\Delta} \phi
    \nonumber\\
    &&\qquad\quad
    - \bar{N} \sqrt{\bar{q}_{xx}}\; \bar{q}_{\vartheta \vartheta} \frac{\partial V}{\partial\phi} \Bigg]
    + \partial_x(\bar{N}^x P_\varphi)\,.
\end{eqnarray}
In Schwarzschild coordinates, such that
\begin{eqnarray}
    &&\bar{N}=\sqrt{1-\frac{2M}{x}} \quad,\quad \bar{q}^{xx}=1-\frac{2M}{x}\,,
    \nonumber\\
    &&\bar{N}^x = 0 \quad,\quad \bar{q}_{\vartheta\vartheta}=x^2\,,
\end{eqnarray}
for a vacuum background, the equations of motion
can be rewritten as the single second order differential equation
\begin{widetext}
\begin{eqnarray}
    \ddot{\phi} &=&
    \frac{1}{x^2} \Bigg[\partial_x \left( x^2 \bar{N} \sqrt{\bar{q}^{xx}}\; (\partial_x\phi)\right)
    + \bar{N} \sqrt{\bar{q}_{xx}} \left(\tilde{\Delta} \phi
    - x^2 \frac{\partial V}{\partial\phi}\right) \Bigg]
    \nonumber\\
    &=&
    \left(1-\frac{2M}{x}\right) \Bigg[
    \frac{2}{x}\left(1-\frac{M}{x}\right) (\partial_x\phi)
    + \left(1-\frac{2M}{x}\right) \partial_x^2\phi
    + \Delta \phi
    -\frac{\partial V}{\partial\phi}
    \Bigg]\,,
\end{eqnarray}
\end{widetext}
which is equivalent to the Klein--Gordon equation (\ref{eq:KG on Schwarzschild}), and hence the same results for quasinormal modes follow.

\section{Spherically symmetric emergent modified gravity}
\label{sec:Spherical emergent modified gravity}

The emergent modified gravity framework, with its systematic approach, provides a reliable method for deriving modified gravitational and matter constraints. It ensures a consistent spacetime structure and offers an expression for the compatible spacetime metric in terms of phase-space degrees of freedom. As we will demonstrate, this framework allows for the explicit derivation of a general version of quasinormal mode equations with a variety of modifications.

\subsection{Vacuum background}
\label{sec:Vacuum background}

We begin with the vacuum backgrounds of spacetime, for which we assume spherical symmetry. The background itself is subject to different modifications, to which scalar matter can be coupled in different ways.

\subsubsection{Canonical system}

Following Refs.~\cite{bojowald2000symmetry,bojowald2004spherically}, we consider the vacuum spherically symmetric theory with canonical gravitational variables $\left\{K_\varphi\left(x\right),E^\varphi\left(y\right)\right\}=\delta\left(x-y\right)$ and $\left\{K_x\left(x\right),E^x\left(y\right)\right\}=\delta\left(x-y\right)$.
In the classical theory, the momenta $E^x$ and $E^\varphi$ are components of the densitized triad, while the configuration variables are directly related to the extrinsic-curvature components ${\cal K}_\varphi =  K_\varphi$ and ${\cal K}_x = 2 K_x$.
Without loss of generality, we will assume that both $E^{\varphi}$ and $E^x$ are positive.
The diffeomorphism constraint is given by
\begin{equation}
    H_x = E^\varphi K_\varphi'
    - K_x (E^x)'\,,
    \label{eq:Diffeomorphism constraint - spherical symmetry}
\end{equation}
and remains unmodified in EMG.

According to EMG \cite{EMGcov}, the most general, covariant Hamiltonian constraint in a spherically symmetric vacuum, up to second-order derivatives and quadratic in first-order derivative terms, is given by equation (\ref{eq:Hamiltonian constraint - modified - non-periodic}) in the Appendix~\ref{app:Spherical emergent modified gravity}.
Here, we will consider the special class of constraints given by
\begin{widetext}
\begin{eqnarray}
    \label{eq:Hamiltonian constraint - modified - non-periodic - special}
    \tilde{H}_{\rm grav}
    \!\!&=&\!\! - \chi \frac{\sqrt{E^x}}{2} \bigg[ E^\varphi \bigg( \frac{1}{E^x}
    + \left(\frac{1}{E^x} - 2 \frac{\partial \ln \lambda^2}{\partial E^x}\right) \frac{\sin^2 \left(\lambda K_\varphi\right)}{\lambda^2}
    + 4 \left(\frac{K_x}{E^\varphi} + \frac{K_\varphi}{2} \frac{\partial \ln \lambda^2}{\partial E^x} \right) \frac{\sin (2 \lambda K_\varphi)}{2 \lambda}
    \bigg)
    \\
    &&
    + \frac{((E^x)')^2}{E^\varphi} \bigg(
    \left( \frac{K_x}{E^\varphi} + \frac{K_\varphi}{2} \frac{\partial \ln \lambda^2}{\partial E^x} \right) \lambda^2 \frac{\sin \left(2 \lambda K_\varphi \right)}{2 \lambda}
    - \frac{\cos^2 \left( \lambda K_\varphi \right)}{4 E^x} \bigg)
    + \left(\frac{(E^x)' (E^\varphi)'}{(E^\varphi)^2}
    - \frac{(E^x)''}{E^\varphi}\right) \cos^2 \left( \lambda K_\varphi \right)
    \bigg]
    \,,\nonumber
\end{eqnarray}
\end{widetext}
with the associated structure function
\begin{eqnarray}
    \tilde{q}^{x x} &=&
    \chi^2 \left( 1
    + \lambda^2 \left( \frac{(E^x)'}{2 E^\varphi} \right)^2
    \right)
    \cos^2 \left( \lambda K_\varphi \right)
    \frac{E^x}{(E^\varphi)^2}
    \nonumber\\
    &=:& \beta \frac{E^x}{(E^\varphi)^2}
    \,,
    \label{eq:Structure function - modified - non-periodic - special}
\end{eqnarray}
where, $\lambda$ is an undetermined function of $E^x$, while $\chi$ is a constant.
The classical constraint is recovered in the limit $\chi\to 1$, $\lambda\to 0$.
The emergent spacetime metric is given by
\begin{equation}\label{eq:Emergent metric}
    {\rm d} s^2 = - N^2 {\rm d} t^2
    + \tilde{q}_{xx} \left({\rm d} x+N^x {\rm d} t\right)^2
    + E^x {\rm d} \Omega^2\,,
\end{equation}
where $\tilde{q}_{xx}=1/\tilde{q}^{xx}$.

\subsubsection{Schwarzschild coordinates}
\label{Sec:Schwarzschild coordinates}

In the Schwarschild gauge~\eqref{eq:Schwarzschild}:
\begin{equation}
    \bar{E}^x=x^2\quad,\quad \bar{N}^x=0\,,
\end{equation}
Hamilton's equations of motion determine the rest of the background gravitational field variables, given by
\begin{eqnarray}
    &&
    \bar{K}_x=\bar{K}_\varphi=0\quad,\quad
    \bar{E}^\varphi=\frac{x}{\sqrt{1-2 M/x}}
    \quad,\nonumber
    \\
    &&
    \bar{N} = \frac{1}{\alpha\chi} \sqrt{1-\frac{2 M}{x}}
    \,,
\end{eqnarray}
where the overline in the variables denote that they are the background gravitational values to be used in the following sections, $\alpha$ is an integration constant that is equivalent to a constant rescaling of the time coordinate.
We choose $\alpha=1/\chi$, for simplicity, and obtain the background spacetime
\begin{widetext}
\begin{eqnarray}
    {\rm d} s^2 &=&
    - \left(1 - \frac{2 M}{x}\right) {\rm d} t^2
    + \beta^{-1} \left(1 - \frac{2 M}{x}\right)^{-1} {\rm d} x^2
    + x^2 {\rm d} \Omega^2
    \nonumber\\
    &=&
    - \left(1 - \frac{2 M}{x}\right) {\rm d} t^2
    + \left( 1 + \lambda^2 \left( 1 - \frac{2 M}{x} \right)
    \right)^{-1} \left(1 - \frac{2 M}{x}\right)^{-1} \frac{{\rm d} x^2}{\chi^2}
    + x^2 {\rm d} \Omega^2
    \,.
    \label{eq:Spacetime metric - modified - Schwarzschild}
\end{eqnarray}
\end{widetext}
In this case, while the equations of motion have solutions identical to their classical form, the space-time geometry is modified because an emergent inverse spatial metric (\ref{eq:Structure function - modified - non-periodic - special}), but not the classical $E^x/(E^{\varphi})^2$, is compatible with the tensor transformation law.
Notice that this line element is not of the standard form $g_{xx}=1/g_{tt}$ such as (\ref{eq:Schwarzschild}) typical of GR and other modified gravity theories.

If we restrict ourselves to asymptotically constant $\lambda$, then the constant $\chi$ may be used to recover an asymptotically flat spacetime by the simple choice $\chi^2=1/\left(1+\lambda_\infty^2\right)$, where $\lambda_\infty=\lim_{x\to\infty} \lambda(x)$.
The two cases of interest are given by constant $\lambda=\bar{\lambda}$ with $\chi^2=1/(1+\bar{\lambda}^2)$ and decreasing $\lambda^2=\Delta/x^2$ with $\chi=1$, which we will use in the following sections.

\begin{figure}[!htb]
    \centering
    \includegraphics[trim=5.4cm 0cm 5.6cm 0cm,clip=true,width=\columnwidth]{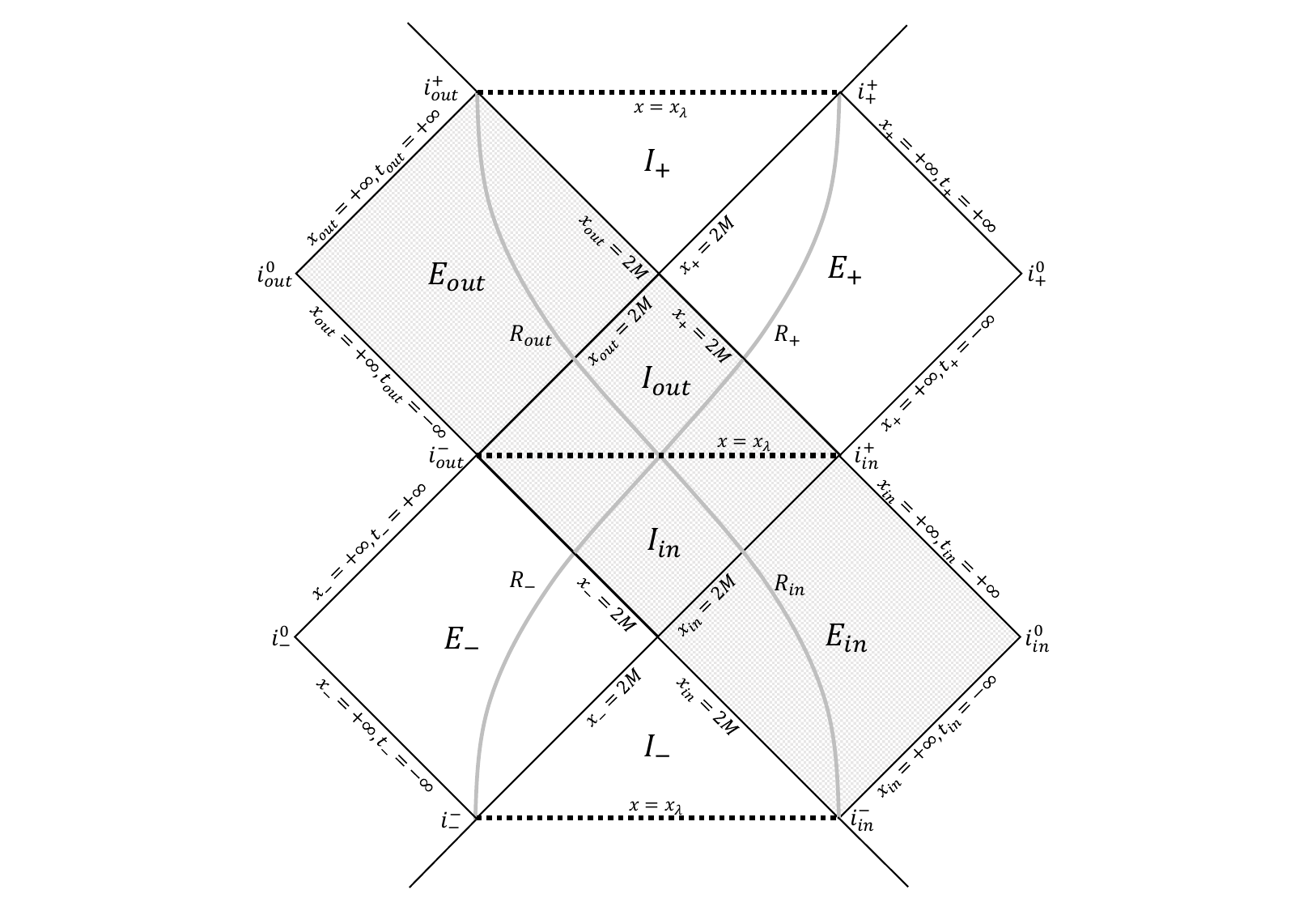}
    \caption{Maximal extension of the vacuum solution. This case has only one solution for the coordinate $x_\lambda$ of the hypersurface of reflection symmetry.
    The region $E_{\rm in}\cup I_{\rm in}\cup I_{\rm out}\cup E_{\rm out}$, corresponding to the shaded area, is the wormhole solution obtained by sewing two interior Schwarzschild regions by the reflection symmetry surface $x=x_\lambda$ using the special gauge that can cross this surface.
    The gray lines $R_{i}$ denote geodesics falling from (to) the remote past (far future). Credit: Ref.~\cite{ELBH}.}
    \label{fig:Holonomy_KS_Vacuum_Wormhole-Periodic}
  \end{figure}

As is discussed in much detail in \cite{ELBH}, the spacetime described by (\ref{eq:Spacetime metric - modified - Schwarzschild}) retains the classical coordinate position of the horizon at $x=2M=:x_{\rm H}$. However, it has a second coordinate singularity at $x=x_\lambda$ defined by the solution to
\begin{equation}
1 + \lambda(x_\lambda)^2 \left( 1 - \frac{2 M}{x_\lambda} \right)=0\,.
\end{equation}
The surface $x=x_\lambda$ is one of reflection symmetry and it implies a minimum-radius.
The global spacetime, therefore, has the structure of an interuniversal wormhole connecting the black hole geometry to a white hole by the common minimum radius surface, which is nonsingular. See Fig.~\ref{fig:Holonomy_KS_Vacuum_Wormhole-Periodic} for the conformal diagram of such a spacetime~\cite{ELBH}.
Further details can be found in the Appendix~\ref{app:Near minimum radius surface}.

\subsubsection{Tortoise coordinates}

We define the tortoise coordinate $x^*$ as 
\begin{eqnarray}\label{eq:Tortoise coord}
    {\rm d} x^* = \frac{{\rm d}x}{\sqrt{\bar{N}^2 \bar{\tilde{q}}^{xx}}}
    \,.
\end{eqnarray}
In this coordinate, the spacetime metric (\ref{eq:Spacetime metric - modified - Schwarzschild}) becomes
\begin{eqnarray}
    {\rm d} s^2 =
    \left(1 - \frac{2 M}{x}\right) \left[ - {\rm d} t^2 + ({\rm d} x^*)^2 \right]
    + x^2 {\rm d} \Omega^2
    \,,
    \label{eq:Spacetime metric - modified - Tortoise}
\end{eqnarray}
where we leave $x$ as an implicit function of $x^*$.

In the asymptotic region $x\to+\infty$, or $x^*\to +\infty$, we may use the expansion $M/x\ll1$, such that to leading order we get
\begin{equation}\label{eq:Tortoise asymptotic}
    x^* \approx x+2M \left(1
    + \frac{1}{2} \frac{\lambda_\infty^2}{1+\lambda_\infty^2}\right) \ln x
    \quad,\quad
    x^*\to+\infty\,,
\end{equation}
where we used $\chi=1/\sqrt{1+\lambda_\infty^2}$.
On the other hand, near the horizon $x=2M+r$, or $x^*\to-\infty$, we may use the expansion $r/(2M)\ll1$, such that to leading order we get
\begin{eqnarray}\label{eq:Tortoise horizon}
    x^* &\approx& \frac{2M}{\chi} \ln (x-2M) 
    - \frac{\lambda_{\rm H}^2}{2 \chi} (x-2M)
    \nonumber\\
    &\approx& \frac{2M}{\chi} \ln (x-2M)
    \quad,\quad
    x^*\to-\infty
    \,,
\end{eqnarray}
where $\lambda_{\rm H}=\lambda(2M)$.

A detailed expansion near the minimum-radius surface can be found in the Appendix~\ref{app:Near minimum radius surface}, which will be useful for the computation of high frequency modes in Section~\ref{sec:High frequency modes}.

\subsection{Scalar matter coupling}
\label{sec:Perturbation theory}

In all cases of this work we consider a scalar matter coupling in spherical symmetry such that the full diffeomorphism constraint takes the form
\begin{equation}
    H_x = E^\varphi K_\varphi' - K_x (E^x)' + P_\phi \phi'
    \,,
\end{equation}
and the Hamiltonian constraint takes the form
\begin{eqnarray}\label{eq:Ham const matter general}
    \tilde{H} \left(E^x,K_x,K_\varphi, E^\varphi, \phi,P_\phi\right)
    = \tilde{H}_{\rm grav}\left(E^x,K_x,K_\varphi, E^\varphi\right)&&
    \nonumber\\
    + \tilde{H}_{\rm scalar}\left(E^x,K_x,K_\varphi, E^\varphi, \phi,P_\phi\right)\quad&&
\end{eqnarray}
where $\tilde{H}_{\rm grav}$ is the vacuum constraint, while the scalar field variables appear only in $\tilde{H}_{\rm scalar}$ with quadratic dependence.

Furthermore, we show in Appendix~\ref{sec:Perturbation theory} that, in the special case of a vacuum background and $\tilde{H}_{\rm scalar}$ with quadratic dependence on the scalar field variables, there is a well-defined perturbative treatment such that backreaction can be neglected in the appropriate regimes. This proof is essential for the results in the following Sections. In the following, we set the gravitational variables equal to the barred functions defined in Subsection~\ref{sec:Vacuum background} to acquire their background values.

\subsection{Minimal coupling of nonspherical scalar matter}

We define minimal coupling in the EMG sense as simply replacing the classical metric with the emergent one in the equations of motion of matter.
In particular, the KG equation on a curved, emergent metric is given by
\begin{eqnarray}\label{eq:KG minimal - EMG}
    \tilde{g}^{\mu\nu} \nabla_\mu \nabla_\nu \phi 
    - \frac{\partial V}{\partial\phi} = 0\,,
\end{eqnarray}
which has been shown to have anomaly-free constraint brackets and to respect covariance of both the spacetime and the matter field in spherical symmetry \cite{EMGscalar}.
As explained in Section~\ref{sec:Classical scalar QNMs}, the non-spherical contributions of the matter field will be encoded in the potential $V$.

To obtain the QNMs equation we shall simply follow the procedure of the classical case with a different background metric.
The equation (\ref{eq:KG minimal - EMG}) can be rewritten as \cite{Das_2005}:
\begin{eqnarray}
\label{eq:KG minimal - EMG01}
    \frac{\partial^2 \tilde{u}_{lm}}{(\partial x^*)^2} + \left[\omega^2-V_l\right] \tilde{u}_{lm}
    = 0
\end{eqnarray}
where $x^*$ is the tortoise coordinate defined by (\ref{eq:Tortoise coord}), and
\begin{eqnarray}\label{eq:Minimal pot}
    V_l \!\!&=&\!\! \frac{l (l+1)}{x^2} \bar{N}^2
    + \frac{\sqrt{\bar{N}^2 \bar{\tilde{q}}^{xx}}}{x} \frac{\partial (\sqrt{\bar{N}^2 \bar{\tilde{q}}^{xx}})}{\partial x}
    \nonumber\\
    \!\!&=&\!\!
    \left(1-\frac{2M}{x}\right) \Bigg[ \frac{l (l+1)}{x^2}
    + \frac{\chi^2}{x} \Bigg[ \frac{2M}{x^2}
    \\
    \!\!&&\!\!
    + \lambda^2 \left(1-\frac{2M}{x}\right) \left(\frac{3M}{x^2} + \frac{1}{2} \frac{\partial \ln \lambda^2}{\partial x}\left(1-\frac{2M}{x}\right)\right)
    \Bigg]\Bigg]
    \,.\nonumber
\end{eqnarray}
It is important to note that the effective potential vanishes at $x\to 2M$ and $x\to +\infty$ for $\lambda(x)$ that do not increase for large $x$. The boundary conditions of $\tilde{u}_{lm}$ therefore closely resemble the classical ones \textemdash\, in fact, as we will see in Section~\ref{sec:Quasinormal modes spectra}, the only deviation is by the constant $\chi$, which is unity for asymptotically vanishing $\lambda$.

Cases of interest are $\lambda=\bar{\lambda}$ (constant) with $\chi^2=1/(1+\bar{\lambda}^2)$ and $\lambda^2=\Delta/x^2$ with $\chi=1$ which give, respectively,
\begin{eqnarray}
    V_l^{(\bar{\lambda})}
    \!\!&=&\!\! \frac{1}{x^2} \left(1-\frac{2M}{x}\right) \left[ l (l+1)
    + \frac{4 M + x_{\bar \lambda}}{2 x}
    - \frac{3M x_{\bar \lambda}}{x^2} \right]
    \nonumber\\
    \!\!&=&\!\! V_l^{({\rm Sch})}
    + \frac{x_{\bar \lambda}}{2x^3} \left(1-\frac{2M}{x}\right) \left( 1 - \frac{6M}{x} \right)
    \,,\\
    V_l^{(\Delta)}
    \!\!&=&\!\!
    \left(1-\frac{2M}{x}\right) \bigg[ \frac{l (l+1)}{x^2} + \frac{2M}{x^3}
    \nonumber\\
    &&\qquad\qquad
    - \frac{\Delta}{x^4} \left(1-\frac{2M}{x}\right) \left(1-\frac{5M}{x}\right)
    \bigg]
    \nonumber\\
    \!\!&=&\!\!
    V_l^{({\rm Sch})}
    - \frac{\Delta}{x^4} \left(1-\frac{2M}{x}\right)^2 \left(1-\frac{5M}{x}\right)
    \,,
\end{eqnarray}
where $V_l^{({\rm Sch})}$ is the classical potential and
\begin{equation}
    x_{\bar \lambda}=2M \frac{\bar{\lambda}^2}{1+\bar{\lambda}^2}
\end{equation}
is the minimum radius in the case of constant $\lambda=\bar{\lambda}$. Interestingly, the classical potential is also obtained for $\lambda=0$ and $\lambda(x)\propto (1-2M/x)^{-3/2}$. However, the latter choice leads to divergence at the horizon.

There are important sign differences between the two schemes.
Compared to the Schwarzschild potential, the barrier will be weakened at $2M<x<6M$ and enhanced at $x>6M$ for constant $\lambda$. However, for a decreasing function, it will be enhanced at $2M<x<5M$ and weakened at $x>5M$.

\begin{figure}[ht]
        \centering
        \includegraphics[trim=1.4cm 0cm 0cm 0cm,clip=true,width=0.9\columnwidth]{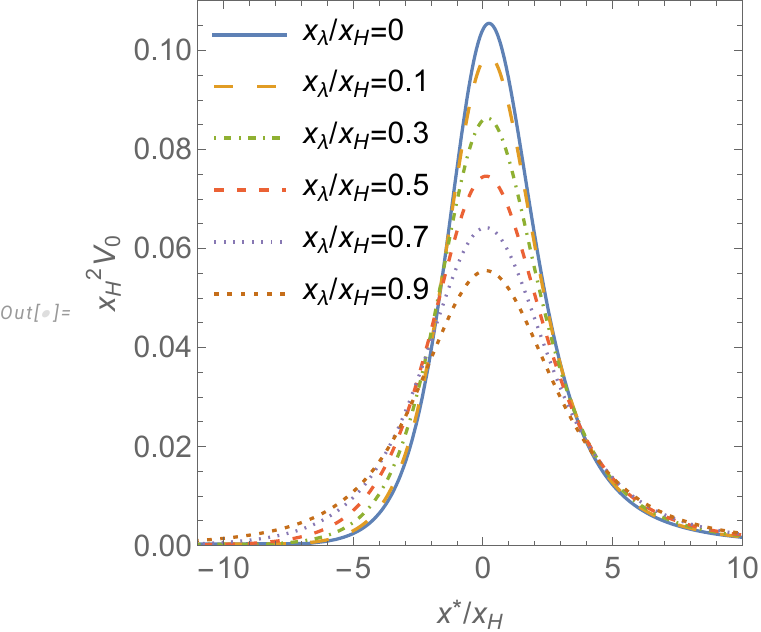}
    %
    \caption{Effective potential $V^{(\bar{\lambda})}_l$ as a function of $x^*/x_H$ 
    with constant parameter $\lambda=\bar{\lambda}$ and $l=0$ for the minimal coupling. The integration for $x^*$ can be made analytically.}
    \label{fig:Mpotconst}
\end{figure}

The effective potential for minimal coupling with constant $\lambda$, shown in Fig.~\ref{fig:Mpotconst}, was previously studied in \cite{Fu_2023,Moreira_2023} and more recently in \cite{Gingrich_2024}.
In this case, the integration (\ref{eq:Tortoise coord}) for $x^*$ can be performed explicitly, obtaining
\begin{widetext}
\begin{equation}\label{eq:x* - holonomy const}
    x_*
    =
    2 M \sqrt{1+\bar{\lambda}^2} \ln \left( c_0
    \left(\frac{1+\sqrt{1 - x_{\bar{\lambda}}/x}}{1-\sqrt{1 - x_{\bar{\lambda}}}/x}\right)^{\frac{1 + x_{\bar{\lambda}}/(4M)}{\sqrt{1+\bar{\lambda}^2}}}
    \left|\frac{\sqrt{1+\bar{\lambda}^2} \sqrt{1 - x_{\bar{\lambda}}/x}-1}{\sqrt{1+\bar{\lambda}^2} \sqrt{1 - x_{\bar{\lambda}}/x}+1}\right| 
    e^{\frac{x}{2M} \frac{\sqrt{1-x_{\bar{\lambda}}/x}}{\sqrt{1+\bar{\lambda}^2}}}
    \right)
\end{equation}
\end{widetext}
where $c_0$ is the integration constant that  in the following will be set equal to one, $c_0=1$, for the sake of simplicity and reproducibility of our results. A different value of $c_0$ would simply displace $x^*$ by a constant but does not affect the system otherwise.

\begin{figure}[ht]
        \centering
        \includegraphics[trim=1.45cm 0cm 0cm 0cm,clip=true,width=0.9\columnwidth]{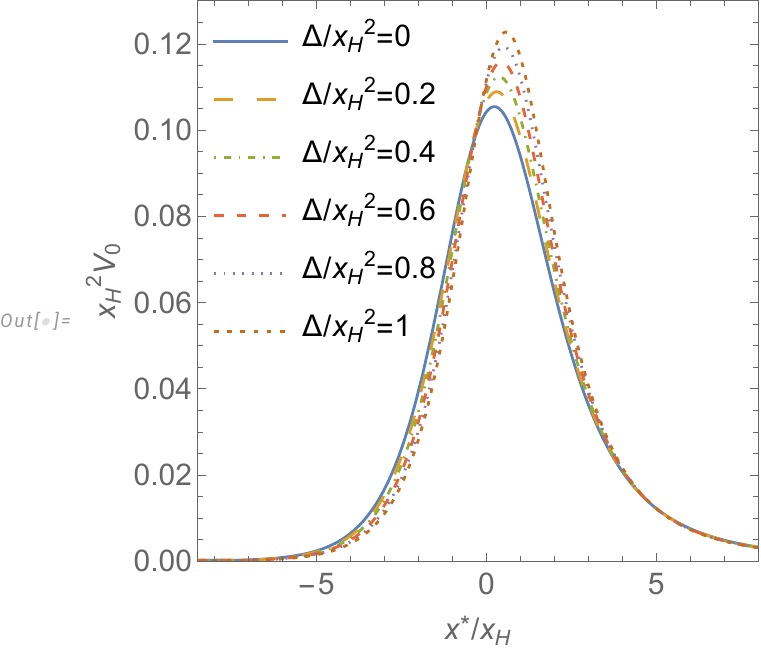}
    %
    \caption{Effective potential $V^{(\Delta)}_l$ as a function of $x^*/x_H$
    with decreasing $\lambda\propto 1/x$ and $l=0$ for the minimal coupling.}
    \label{fig:Mpotmubar}
\end{figure}

Similarly, the effective potential of the minimal coupling with decreasing $\lambda\propto 1/x$ is shown in Fig.~\ref{fig:Mpotmubar}, which was not included in \cite{Fu_2023,Moreira_2023} because the dynamical solution had not been known until recently \cite{ELBH}.
Unlike the case of a constant $\lambda$, here the potential barrier increases as a function of $\lambda$.
In this case, the integration for $x^*$ must be done numerically and we fix $x^*(x=1000)=x^*_{\rm (classical)}(x=1000)$ for unambiguous and reproducible results.

The potential $V_l$ represents a barrier that the perturbations must overcome to propagate outward.
The higher the potential barrier, the less likely it is that quasinormal modes will be produced, implying smaller amplitudes.
Therefore, since for a  constant $\lambda$, Fig.~\ref{fig:Mpotconst}, the peak of the potential barrier decreases as $\bar{\lambda}$ increases, it implies that the perturbations can escape the region near the black hole more easily than in the classical case.
On the other hand, for a decreasing parameter, Fig.~\ref{fig:Mpotmubar}, the peak of the potential barrier increases as $\Delta$ increases, and we obtain the opposite result: it becomes more difficult for perturbations to escape the region near the black hole than in the classical case.

\begin{figure}[ht]
    \centering
    \includegraphics[trim=1.4cm 0cm 0cm 0cm,clip=true,width=0.9\columnwidth]{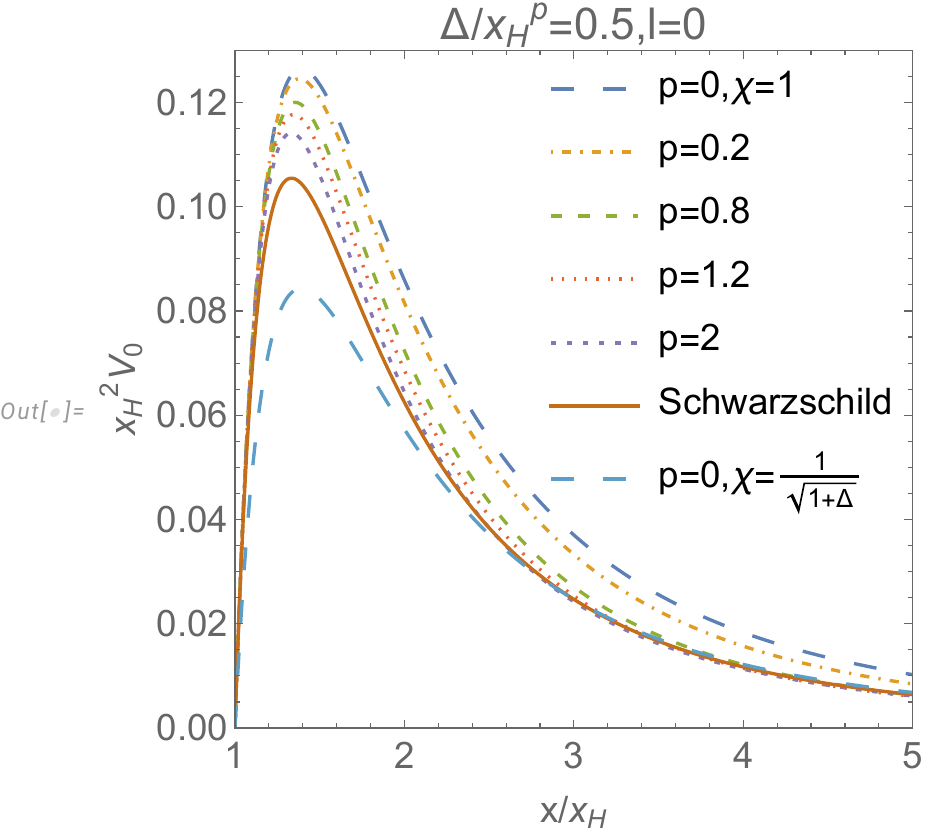}
    \caption{Effective potential $V^{(\Delta)}_l$ as a function of $x/x_H$ with a power law $\lambda^2=\Delta/x^p$ and $l=0$ for the minimal coupling. The power parameter $p$ is varied from $p=0$ to $p=2$ while the proportionality constant is fixed to $\Delta/x_{\rm H}^p=0.5$.}
    \label{fig:Mpotpower}
\end{figure}

That the peak of the potential barrier is lower in the case of constant $\lambda=\bar{\lambda}$ compared to the decreasing $\lambda\propto1/x$ is a direct consequence of the choice $\chi=1/\sqrt{1+\bar{\lambda}^2}$ in the former.
Indeed, if $\chi=1$ is chosen, which is the choice that makes the decreasing case asymptotically flat, then the barrier is higher than the classical one and can even be higher than the decreasing $\lambda$ case.
See Fig.~\ref{fig:Mpotpower} for an illustration of this.
The choice $\chi=1/\sqrt{1+\bar{\lambda}^2}$ for constant $\lambda=\bar{\lambda}$ was motivated by the recovery of an asymptotically flat spacetime.
However, as we have just argued, the precise value of the constant $\chi$ can have important physical effects, including the size of the potential barrier of QNMs.
In fitting the theoretical results of QNMs with observational data, we should independently consider the variation of both parameters, $ \lambda $ and $ \chi $.
In the following sections, however, we will restrict our analysis to $\chi$ that recovers an asymptotically flat spacetime.

\subsection{Nonminimal coupling of nonspherical scalar matter}
\label{sec:Nonminimal coupling}

The scalar matter can be coupled in more forms such that its equation of motion need not resemble the Klein-Gordon equation even while it remains covariant \cite{EMGscalar}.
One such realization  is given by the following spherically symmetric Hamiltonian constraint (a more general case is discussed in Appendix~\ref{app:Scalar canonical system} with the general constraint (\ref{eq:Hamiltonian constraint - DF - scalar polymerization - roots}))
\begin{widetext}
\begin{eqnarray}
    \tilde{H} &=&
    - \chi \frac{\sqrt{E^x}}{2} \bigg[ E^\varphi \bigg( \frac{1}{E^x}
    + \left(\frac{1}{E^x} - 2 \frac{\partial \ln \lambda^2}{\partial E^x}\right) \frac{\sin^2 \left(\lambda K_\varphi\right)}{\lambda^2}
    + 4 \left(\frac{K_x}{E^\varphi} + \frac{K_\varphi}{2} \frac{\partial \ln \lambda^2}{\partial E^x} \right) \frac{\sin (2 \lambda K_\varphi)}{2 \lambda}
    \bigg)
    \nonumber\\
    &&
    + \frac{((E^x)')^2}{E^\varphi} \bigg(
    \left( \frac{K_x}{E^\varphi} + \frac{K_\varphi}{2} \frac{\partial \ln \lambda^2}{\partial E^x} \right) \lambda^2 \frac{\sin \left(2 \lambda K_\varphi \right)}{2 \lambda}
    - \frac{\cos^2 \left( \lambda K_\varphi \right)}{4 E^x} \bigg)
    + \left(\frac{(E^x)' (E^\varphi)'}{(E^\varphi)^2}
    - \frac{(E^x)''}{E^\varphi}\right) \cos^2 \left( \lambda K_\varphi \right)
    \bigg]
    \nonumber\\
    &&
    - \chi \frac{\sqrt{E^x}}{2} \Bigg[ - \frac{P_\phi{}^2}{E^\varphi} \frac{1}{E^x} \left(1+\lambda^2 \left(\frac{(E^x)'}{2E^\varphi}\right)^2\right) \cos^2 (\lambda K_\varphi)
    - \frac{E^x}{E^\varphi} (\phi')^2
    - 2 E^\varphi V
    \Bigg]
    \nonumber\\
    &=:& \tilde{H}_{\rm grav}
    + \tilde{H}_{\rm scalar}
    \,,
    \label{eq:Hamiltonian constraint - DF - nonperiodic - simple}
\end{eqnarray}
\end{widetext}
where $V$ is an undetermined function of $E^x$ and $\phi$ representing the scalar potential.
Here, $\tilde{H}_{\rm grav}$ is precisely the vacuum constraint (\ref{eq:Hamiltonian constraint - modified - non-periodic - special}) and $\tilde{H}_{\rm scalar}$ is quadratic in the matter variables assuming the potential is quadratic too.
The presence of non-minimal coupling of the scalar field to gravity can be seen by noting that only the $P_{\phi}$-term contains a factor of the emergent spatial metric (\ref{eq:Structure function - DF - roots}) with our choice of parameters, while the $\phi'$-term and the scalar potential $V$ appears only in combination with the classical densitized-triad fields.

We now define $\tilde{H}_{\rm grav}$ as the constraint of the gravitational background for a scenario without backreaction, and hence all equations of motion for the gravitational field are generated by $\tilde{H}_{\rm grav}$. The constraint surface, given by $\tilde{H}_{\rm grav}=0$ and $H^{\rm grav}_x=0$, implies a background solution equal to the one obtained in Subsection~\ref{sec:Vacuum background}.
The Hamiltonian for the scalar matter is then given by $\tilde{H}_{\rm scalar}[\bar{N}]$.
See Appendix~\ref{sec:Perturbation theory} for details of our perturbation theory.

The equations of motion for the scalar matter in this case are
\begin{eqnarray}
    \dot{\phi} &=& 
    \chi^{-1} \beta \left(1-\frac{2 M}{x}\right) \frac{P_\phi}{x^2}
    \,,\\
    \dot{P}_\phi
    &=& 
    \chi \Bigg[ \left(
    2 \left(\frac{x^2}{2}-xM\right) \phi'\right)'
    - x^2 \frac{\partial V}{\partial \phi} \Bigg]
    \,.
\end{eqnarray}
These can be combined into the single equation
\begin{eqnarray}
    0 &=&
    - \beta^{-1} \left(1-\frac{2 M}{x}\right)^{-1} \ddot{\phi}
    + \frac{2}{x} \left(1-\frac{M}{x}\right) \phi'
    \nonumber\\
    &&\qquad
    + \left(1-\frac{2M}{x}\right) \phi''
    - \frac{\partial V}{\partial \phi}\,.
\end{eqnarray}
Introducing the non-spherical terms can be done from the Hamiltonian $\tilde{H}_{\rm scalar}$ or directly in the equations of motion by including a 2-dimensional Laplacian of the angular coordinates. The expansion (\ref{eq:Spherical harmonics exp}) in spherical harmonics amounts to using the effective potential (\ref{eq:Effective potential angular momentum}) and results in
\begin{eqnarray}
    0 &=&
    - \beta^{-1} \left(1-\frac{2 M}{x}\right)^{-1} \ddot{\tilde{\phi}}_{lm}
    + \frac{2}{x} \left(1-\frac{M}{x}\right) \partial_x\tilde{\phi}_{lm}
    \nonumber\\
    &&
    + \left(1-\frac{2M}{x}\right) \partial_x^2\tilde{\phi}_{lm}
    - \frac{l(l+1)}{x^2} \tilde{\phi}_{lm}
\end{eqnarray}
After Fourier transformation,
\begin{eqnarray} \label{Fouriert}
    \tilde{\phi}_{lm} (t,x) &=& \frac{u_{lm} (t,x)}{x}\,,\\
    u_{lm} (t,x) &=& \int^\infty_{-\infty} \frac{{\rm d} \omega}{2\pi} \tilde{u}_{lm} (\omega,x) e^{-i\omega t}\,,
\end{eqnarray}
the equation becomes
\begin{eqnarray}\label{eq:Noniminal QNM eq Fourier}
    0 &=&
    \beta \left(1-\frac{2M}{x}\right)^2 \partial_x^2\tilde{u}_{lm}
    \\
    &&
    + \beta \left(1-\frac{2 M}{x}\right) \frac{2M}{x^2} \partial_x\tilde{u}_{lm}
    \nonumber\\
    &&
    + \left[\omega^2
    - \beta \left(1-\frac{2 M}{x}\right) \left(\frac{l(l+1)}{x^2}+\frac{2M}{x^3}\right)\right] \tilde{u}_{lm}
    \,.\nonumber
\end{eqnarray}

This equation can be further simplified by defining the new coordinate $x^*$ as the tortoise coordinate defined by (\ref{eq:Tortoise coord}), such that (\ref{eq:Noniminal QNM eq Fourier}) becomes
\begin{eqnarray}\label{eq:Noniminal QNM eq Fourier x*}
    0 &=&
    \frac{\partial^2 \tilde{u}_{lm}}{(\partial x^*)^2}
    - \frac{\chi^2 \lambda^2}{\sqrt{\beta}} \left(1-\frac{2 M}{x}\right)
    \nonumber\\
    &&
    \qquad\times
    \left( \frac{2 M}{x^2}
    + \left(1-\frac{2 M}{x}\right) \frac{\partial\ln \lambda^2}{\partial x} \right) \frac{\partial \tilde{u}_{lm}}{\partial x^*}
    \nonumber\\
    &&
    + \left[ \omega^2
    - \beta \left(1-\frac{2 M}{x}\right) \left(\frac{l(l+1)}{x^2}+\frac{2M}{x^3}\right) \right] \tilde{u}_{lm}
    \nonumber\\
    &=:&
    \frac{\partial^2 \tilde{u}_{lm}}{(\partial x^*)^2}
    + 2 \zeta \frac{\partial \tilde{u}_{lm}}{\partial x^*}
    + \left[\omega^2 - V_l (x)\right] \tilde{u}_{lm}
\end{eqnarray}
where
\begin{eqnarray}
    \!\!&&\!\!\hspace{-0.5cm}
    V_l (x) = \frac{\beta}{x^2} \left(1-\frac{2 M}{x}\right) \left(l(l+1)+\frac{2M}{x}\right)
    \,,\\
    \!\!&&\!\!\hspace{-0.5cm}
    \zeta = - \frac{\chi^2 \lambda^2}{2 \sqrt{\beta}} \left(1-\frac{2 M}{x}\right) \left( \frac{2 M}{x^2}
    + \left(1-\frac{2 M}{x}\right) \frac{\partial\ln \lambda^2}{\partial x} \right)
    .\label{eq:Damping term - general}
\end{eqnarray}
In contrast to the minimally coupled scenario, here we have an (anti)damping term $\zeta$ and further corrections to the potential $V_{l}$.

\begin{figure}[!htb]
        \centering
        \includegraphics[trim=1.4cm 0cm 0cm 0cm,clip=true,width=0.9\columnwidth]{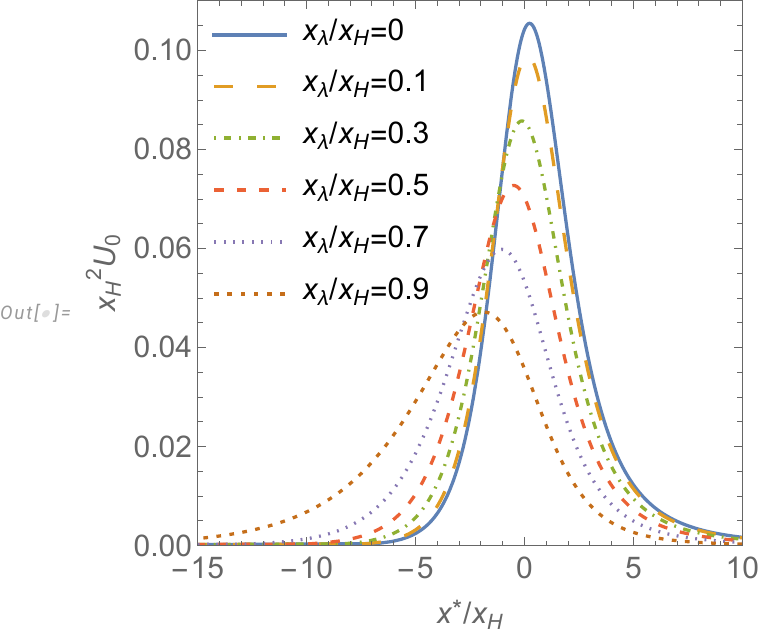}
    %
    \caption{Effective potential $U_l$ as a function of $x^*/x_{\rm H}$
    with constant $\lambda$ and $l=0$ for the nonminimal coupling.}
    \label{fig:NMpotconst}
\end{figure}

Defining
\begin{equation}\label{eq:Fourier-modulation relation}
    \tilde{u}_{lm} (\omega,x^*) = Z(x^*) S_{lm} (\omega,x^*)\,,
\end{equation}
and
\begin{equation} \label{modulation}
    Z (x^*) = \exp \left( - \int{\rm d} x^*\; \zeta \right)
    \,,
\end{equation}
equation (\ref{eq:Noniminal QNM eq Fourier x*}) becomes of Schr\"odinger type,
\begin{equation}\label{eq:QNM equation - Schrodinger}
    \frac{\partial^2 S_{lm}}{(\partial x^*)^2} + \left[\omega^2-U_l(x^*)\right] S_{lm} = 0
    \,,
\end{equation}
with the $\omega$-independent potential
\begin{eqnarray}\label{eq:Potential - Schrodinger}
    U_l = V_l + \zeta^2- \frac{\partial \zeta}{\partial x^*}
    \,.
\end{eqnarray}
This potential stands for arbitrary $\lambda(x)$ and therefore its deviation from the classical potential will depend on the specific choice of this function.
However, as a general feature, the potential does vanish at $x\to 2M$ and $x\to +\infty$, for $\lambda(x)$ that do not increase for large $x$, and therefore the boundary conditions will resemble the classical ones just as it happens with the minimally coupled case.
In Sections~\ref{sec:QNM Constant holonomy function} and \ref{sec:QNM decreasing holonomy function} we present the explicit expressions for the potentials and (anti)damping factors for constant and decreasing modification functions $\lambda(x)$, respectively.
Note that the (anti)damping term has formally disappeared after transforming to equation (\ref{eq:QNM equation - Schrodinger}), but its effects remain in the modulation factor (\ref{modulation}) of the QNM amplitudes.

As can be seen in Fig.~\ref{fig:NMpotconst}, the effective potential of the nonminimal coupling with constant parameter $\bar{\lambda}$ differs significantly from the minimal one shown in Fig.~\ref{fig:Mpotconst}. As the parameter $\bar{\lambda}$ increases, the potential barrier decreases more for nonminimal coupling than for minimal coupling.

\begin{figure}[ht]
        \centering
        \includegraphics[trim=1.3cm 0cm 0cm 0cm,clip=true,width=0.9\columnwidth]{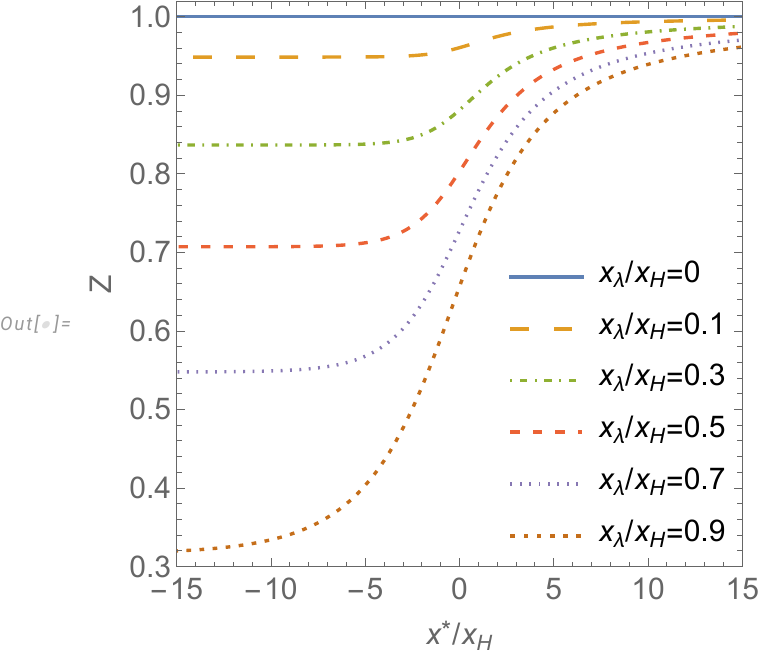}
    \caption{Modulation factor $Z$ as a function of $x^*/x_H$
    where $x_H=2M$, with constant $\lambda$.}
    \label{fig:NMdampconst}
\end{figure}

The (anti)damping factor has a negative sign near the horizon but its integration results in $Z(x^*)<1$. However, this function approaches unity asymptotically for large $x$; see Fig.~\ref{fig:NMdampconst}.

\begin{figure}[ht]
        \centering
        \includegraphics[trim=1.4cm 0cm 0cm 0cm,clip=true,width=1\columnwidth]{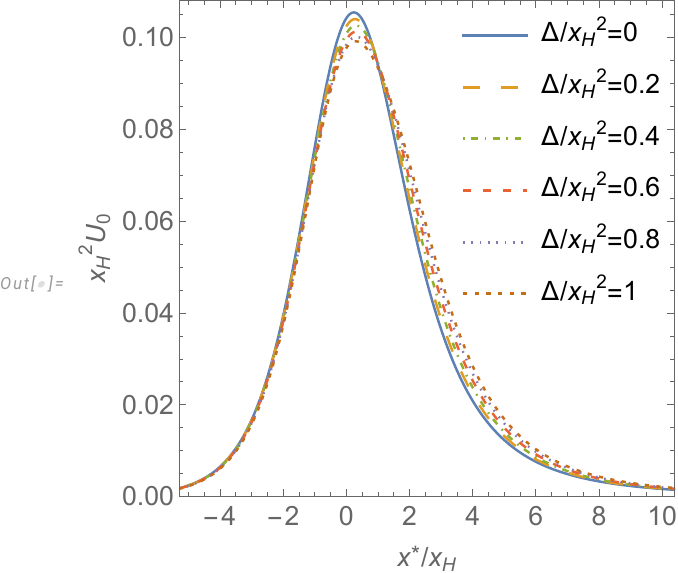}
    \caption{Effective potential $U_l$ as a function of $x^*/x_H$
    with decreasing $\lambda\propto 1/x$ and $l=0$ for the nonminimal coupling.}
    \label{fig:NMpotmubar}
\end{figure}

\begin{figure}[ht]
        \centering
        \includegraphics[trim=1.4cm 0cm 0cm 0cm,clip=true,width=0.9\columnwidth]{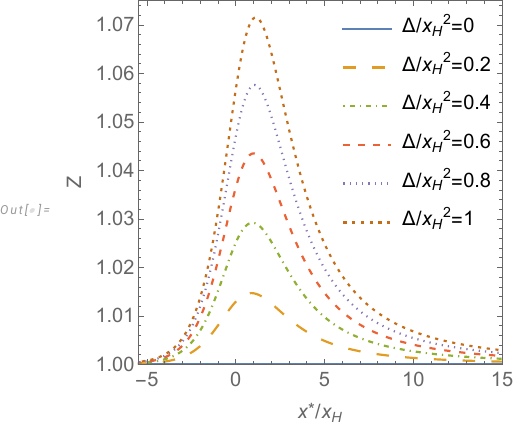}
    \caption{Modulation factor $Z$ as a function of $x^*/x_H$
    where $x_H=2M$, with decreasing $\lambda\propto 1/x$.}
    \label{fig:NMdampmubar}
\end{figure}

Fig.~\ref{fig:NMpotmubar} contains the plot of the effective potential of the nonminimal coupling with decreasing $\lambda\propto 1/x$. Unlike the potential for constant $\lambda$, as seen in Fig.~\ref{fig:NMpotconst}, the new potential does not differ significantly from the classical one.
Furthermore, it differs from the minimally coupled potential, Fig.~\ref{fig:Mpotmubar}, in that the peak of the potential increases in the latter case, while it decreases for nonminimal coupling. The (anti)damping factor has a negative sign near the horizon and, contrary to the constant parameter case, it amplifies the waves. It decreases to unity asymptotically for large $x$;
see Fig.~\ref{fig:NMdampmubar}.

Since the potential barrier decreases with the modification parameter $\lambda$ in both constant and decreasing cases, we conclude that the quasinormal modes are more likely to be produced or tend to have higher amplitudes than in the classical case.
Such amplitudes will be further pumped by the respective (anti)damping factors.
We conclude that the perturbations are generically stronger in nonminimal coupling of EMG than in classical gravity.

\section{Spectra of quasinormal modes}
\label{sec:Quasinormal modes spectra}

Coefficients in the equation for quasinormal modes have shown characteristic features for different cases of emergent modified gravity. As usual, the close relationship to the Schr\"odinger equation of quantum mechanics makes it possible to derive properties of solutions such as the frequency spectrum by standard methods.
A brief review of how to set up the problem, especially the boundary conditions, is presented in Appendix~\ref{Sec:Standard case}.

\subsection{WKB approximation}
\label{sec:WKB approximation}

Our procedure in Subsection~\ref{sec:Nonminimal coupling} showed that the nonminimal equations of motion can be reduced to a Schr\"odinger-type equation with eigenvalue $\omega^2$,
\begin{equation}\label{eq:QNM equation - Schrodinger - WKB}
    \frac{\partial^2 S_{lm}}{(\partial x^*)^2} + \left[\omega^2-U_l(x^*)\right] S_{lm} = 0
    \,,
\end{equation}
where the potential $U_l(x^*)$ is given by (\ref{eq:Potential - Schrodinger}), and $S_{lm}$ is the modulated amplitude (\ref{eq:Fourier-modulation relation}).
The minimally coupled case follows a Schr\"odinger-type equation too given by (\ref{eq:KG minimal - EMG01}), differing from (\ref{eq:QNM equation - Schrodinger - WKB}) only by the replacement of $S_{lm}\to \tilde{u}_{lm}$, since there is no modulation or (anti)damping in the minimal coupling, and $U_l(x^*)\to V_l$ with the latter being the minimally coupled potential (\ref{eq:Minimal pot}). The two cases can, therefore, be solved by standard methods for eigenvalue problems.
In particular, as we showed in the previous section, the minimal and nonminimal potentials have a well-defined peak and vanish both at spatial infinity and at the horizon for well-behaved $\lambda(x)$\textemdash\,in particular, this is true for our cases of interest of constant $\lambda(x)=\bar{\lambda}$ and decreasing $\lambda(x)=\sqrt{\Delta/x^2}$\textemdash\,, indicating that the WKB approximation is well-suited for the system.
Before we delve into the WKB method, it is crucial to first discuss how the boundary conditions may be modified in EMG.

Because the effective potential in EMG vanishes asymptotically and at the horizon\textemdash\, which is valid for both the minimal (\ref{eq:Minimal pot}) and nonminimal (\ref{eq:Potential - Schrodinger}) potentials that do not increase for large $x$\textemdash\,, the solutions to (\ref{eq:QNM equation - Schrodinger - WKB}) in these regions and the boundary conditions are identical to those of the standard case discussed in Subsection~\ref{Sec:Standard case} with the only difference being in the relation $x^*(x)$ which is defined by (\ref{eq:Tortoise coord}) in EMG.
Therefore, using (\ref{eq:Tortoise asymptotic}) and (\ref{eq:Tortoise horizon}), the boundary conditions at the horizon and the asymptotic region, which are respectively given by (\ref{eq:Boundary cond - horizon - standard}) and (\ref{eq:Boundary cond - Inf - standard}) in standard GR, respectively become
\begin{widetext}
\begin{eqnarray}
    \label{eq:Boundary cond - horizon}
    S_{lm} (\omega,x^*) \!\!&\approx&\!\! e^{- i\omega x^*}
    \approx \left(x-2M\right)^{- i x_{\rm H}\omega /\chi}
    \qquad,\quad x^* \to - \infty
    \,,\\
    \label{eq:Boundary cond - Inf}
    S_{lm} (\omega,x^*) \!\!&\approx&\!\! e^{+ i\omega x^*}
    \approx x^{i\omega x_{\rm H} \left(1
    + \chi^2 \lambda_\infty^2 / 2\right)} e^{+ i\omega x}
    \quad,\quad x^* \to + \infty
    \,,
\end{eqnarray}
\end{widetext}
where the signs of the argument of the complex exponentials respectively restrict the solutions to those of inward propagating modes at the horizon, so nothing comes out of the black hole, and outward propagating modes in the asymptotic region, so nothing comes in from outside the universe, as discussed in detail in Appendix~\ref{Sec:Standard case}.
Notice that we recover precisely the classical expressions for both boundary conditions only with an asymptotically vanishing $\lambda(x)$, such that $\lambda_\infty\to0$ and $\chi\to1$.

Having set up a standard eigenvalue problem, we now apply the third order WKB approximation, which improves the accuracy of the resulting complex eigenfrequencies with respect to the first order approximation and are consistent with more complex computational techniques for the classical Schwarzschild black hole \cite{IyerWKB}. 
The resulting eigenvalues are given by
\begin{equation}\label{eq:WKB eigenfrequency}
    \omega^2_{nl} \approx U_{l\, {\rm max}} + \sqrt{-2 U_{l\, {\rm max}}^{(2)}}\; \left[\Lambda
    - i \left(n+\frac{1}{2}\right) \left(\Omega+1\right) \right]
    \,,
\end{equation}
where
\begin{eqnarray}
    \Lambda &=& \frac{1}{\sqrt{-2 U_{l\, {\rm max}}^{(2)}}} \left[ \frac{1}{8} \frac{U_{l\, {\rm max}}^{(4)}}{U_{l\, {\rm max}}^{(2)}} \left(\frac{1}{4} + \left(n+\frac{1}{2}\right)^2\right)
    \right.
    \\
    &&\left.
    \qquad\qquad
    - \frac{1}{288} \left(\frac{U_{l\, {\rm max}}^{(3)}}{U_{l\, {\rm max}}^{(2)}}\right)^2 \left(7+60\left(n+\frac{1}{2}\right)^2\right) \right]
    \nonumber
\end{eqnarray}
and
\begin{eqnarray}
    \Omega \!\!&=&\!\! \frac{1}{-2 U_{l\, {\rm max}}^{(2)}} \Bigg[
    \frac{5}{6912} \left(\frac{U_{l\, {\rm max}}^{(3)}}{U_{l\, {\rm max}}^{(2)}}\right)^4 \left(77+188 \left(n+\frac{1}{2}\right)^2\right)
    \nonumber\\
    \!\!&&\!\!
    - \frac{1}{384} \left(\frac{U_{l\, {\rm max}}^{(3)}}{U_{l\, {\rm max}}^{(2)}}\right)^2 \frac{U_{l\, {\rm max}}^{(4)}}{U_{l\, {\rm max}}^{(2)}} \left(51+100\left(n+\frac{1}{2}\right)^2\right)
    \nonumber\\
    \!\!&&\!\!
    + \frac{1}{2304} \left(\frac{U_{l\, {\rm max}}^{(4)}}{U_{l\, {\rm max}}^{(2)}}\right)^2 \left(67+68\left(n+\frac{1}{2}\right)^2\right)
    \nonumber\\
    \!\!&&\!\!
    + \frac{1}{288} \frac{U_{l\, {\rm max}}^{(3)}}{U_{l\, {\rm max}}^{(2)}} \frac{U_{l\, {\rm max}}^{(5)}}{U_{l\, {\rm max}}^{(2)}} \left(19+28\left(n+\frac{1}{2}\right)^2\right)
    \nonumber\\
    \!\!&&\!\!
    - \frac{1}{288} \frac{U_{l\, {\rm max}}^{(6)}}{U_{l\, {\rm max}}^{(2)}} \left(5+4\left(n+\frac{1}{2}\right)^2\right)
    \Bigg]\,.
\end{eqnarray}
Here, $U_{l\, {\rm max}}$ is the maximum value of the potential $U_{l}$, while $U_{l\, {\rm max}}^{(i)}$ is the $i$th derivative with respect to $x^*$ of the potential evaluated at the point where the latter is maximized. The integers $n=0,1,2,\dots$ and $l=0,1,2,\dots$ label eigenvalues $\omega_{nl}^2$.

As discussed in more detail in Appendix~\ref{Sec:Standard case}, the actual frequencies are given by the square root of the eigenvalues and lead to two branches differing from one another by an overall sign $\omega^\pm=\pm\sqrt{\omega_{nl}^2}$, where the positive branch frequency $\omega^+$ is defined by a positive real value, ${\rm sgn}\left({\rm Re}[\omega^+]\right)=+1$.
The boundary conditions restrict the solutions to only those with ${\rm sgn}\left({\rm Im} [\omega^\pm]\right)=-1$ since the alternative becomes negligible at the asymptotic region and is hence unobservable.
As shown below, for specific values of $\lambda$, the positive branch frequency $\omega^+$ may have a positive imaginary part while the negative branch frequency $\omega^-$ may have a negative imaginary part.
In such a case, it is not just beneficial, but necessary to consider both branches to describe the entire observable spectrum of the QNMs. We define it as $ {\omega_{nl}^{\rm (obs)}}= {\omega_{nl}^+ \cup \omega_{nl}^-| {\rm sgn}\left({\rm Im}[\omega_{nl}^\pm]\right) = -1}$.
We will give examples of this in the following Subsections.
In the following, $S_{lm}$ stands for either branch and will discriminate them with the labels $\pm$ in $S^\pm_{lm}$ and their respective eigenfrequencies $\omega^\pm$ only when there is potential confusion.

\subsection{Constant modification function}
\label{sec:QNM Constant holonomy function}

In this subsection we will focus entirely on the case of constant $\lambda^2=\bar{\lambda}^2$ with $\chi^2=1/(1+\bar{\lambda}^2)$.
The nonminimal potential and the (anti)damping and modulation  factors in this case are given by
\begin{eqnarray}
    V_l (x) \!\!&=&\!\! 
    \frac{1}{x^2} \left(1-\frac{x_{\bar \lambda}}{x}\right) \left(1-\frac{2M}{x}\right) \left(l(l+1)
    + \frac{2M}{x}\right)
    \\
    \zeta \!\!&=&\!\! - \frac{x_{\bar{\lambda}}}{2x^2} \left(1-\frac{x_{\bar \lambda}}{x}\right)^{-1/2} \left(1-\frac{2 M}{x}\right)
    \,,\\
    U_l \!\!&=&\!\! \frac{1}{x^2} \left(1-\frac{2 M}{x}\right)
    \\
    \!\!&&\!\!\times\left( l(l+1) \left(1-\frac{x_{\bar\lambda}}{x}\right)
    - \frac{x_{\bar\lambda}}{x}
    + \frac{2M}{x} \left(1+\frac{x_{\bar\lambda}}{2x}\right)\right)\nonumber
    \,,\\
    Z \!\!&=&\!\! \sqrt{1-\frac{x_\lambda}{x}}
    \,,
\end{eqnarray}
See Figs.~\ref{fig:NMpotconst} and \ref{fig:NMdampconst} for plots of the effective potential and the modulation factor, respectively, for different values of $\bar{\lambda}$.

\begin{figure*}[!ht]
    \begin{subfigure}{.32\textwidth}
        \includegraphics[trim=1.5cm 0cm 0cm 0cm,clip=true,width=1\textwidth]{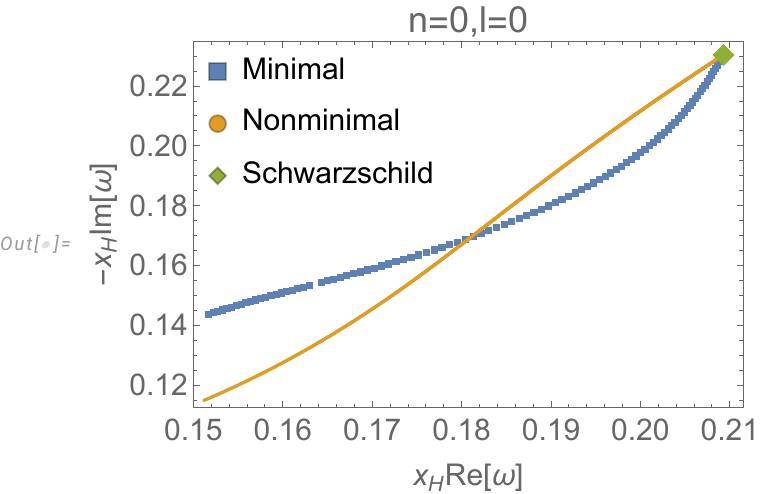}
        \caption{\empty}
        \label{fig:w00const}
    \end{subfigure}
    \begin{subfigure}{.32\textwidth}
        \includegraphics[trim=1.5cm 0cm 0cm 0cm,clip=true,width=1\textwidth]{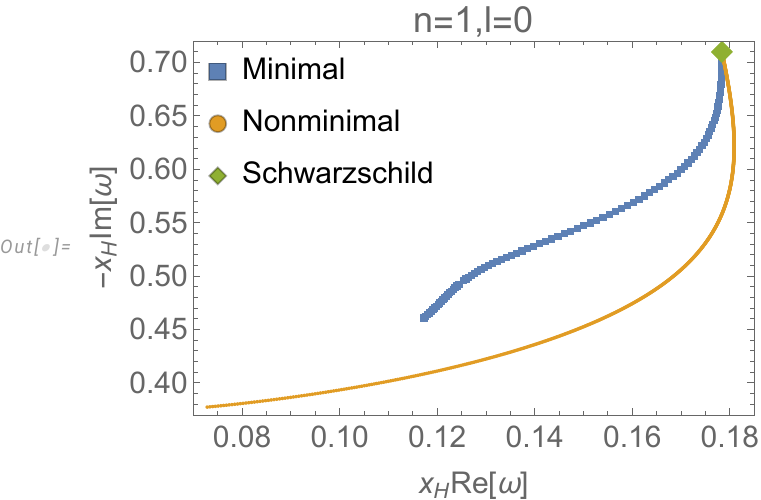}
        \caption{\empty}
        \label{fig:w10const}
    \end{subfigure}
    \begin{subfigure}{.32\textwidth}
        \includegraphics[trim=1.5cm 0cm 0cm 0cm,clip=true,width=1\textwidth]{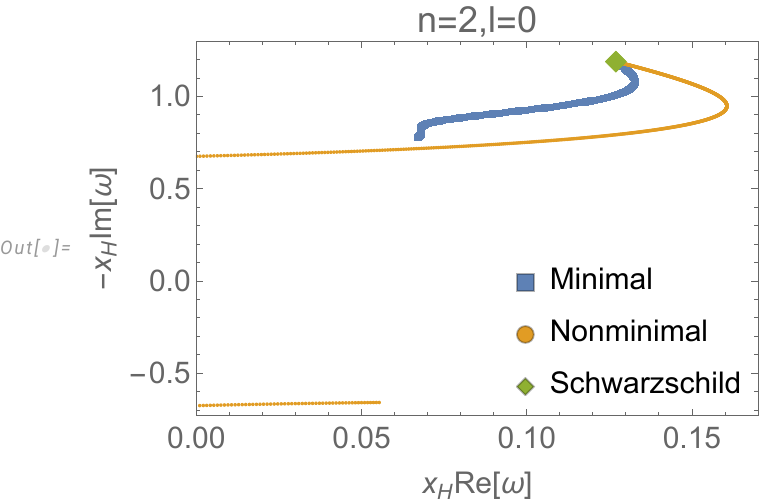}
        \caption{\empty}
        \label{fig:w20const}
    \end{subfigure}
    \begin{subfigure}{.32\textwidth}
        \centering
        \includegraphics[trim=1.5cm 0cm 0cm 0cm,clip=true,width=1\textwidth]{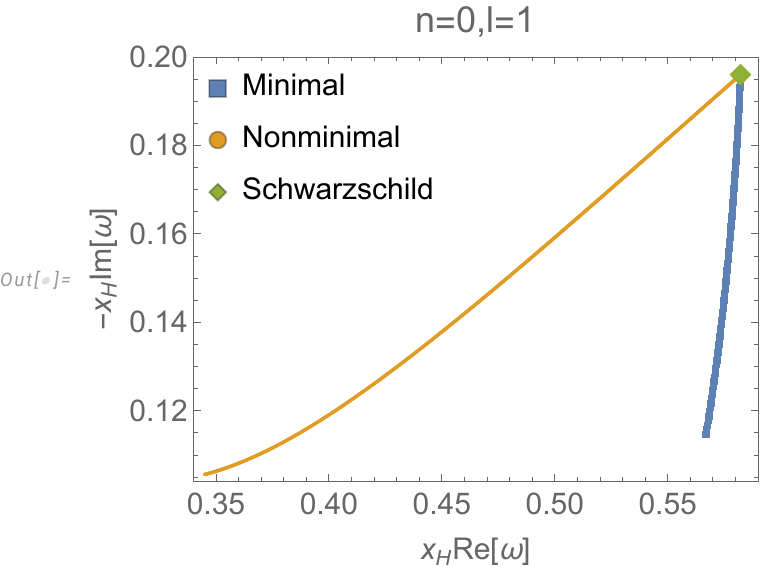}
        \caption{\empty}
        \label{fig:w01const}
    \end{subfigure}%
    \begin{subfigure}{.3325\textwidth}
        \centering
        \includegraphics[trim=1.5cm 0cm 0cm 0cm,clip=true,width=1\textwidth]{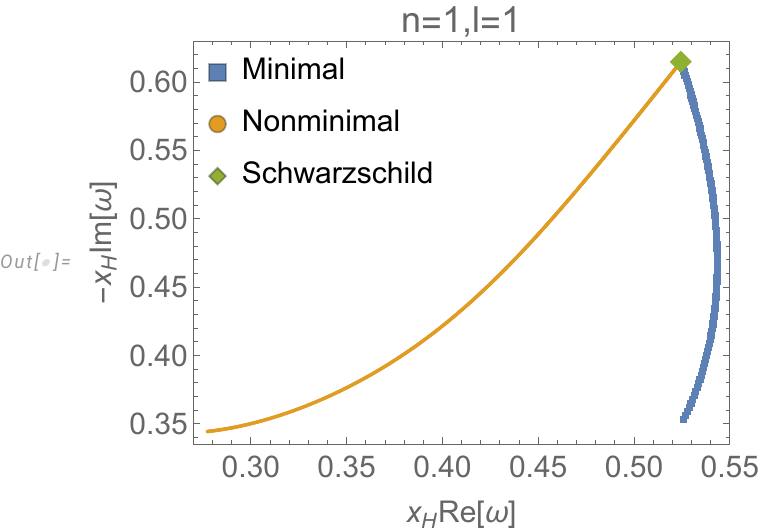}
        \caption{\empty}
        \label{fig:w11const}
    \end{subfigure}
    \begin{subfigure}{.32\textwidth}
        \centering
        \includegraphics[trim=1.5cm 0cm 0cm 0cm,clip=true,width=1\textwidth]{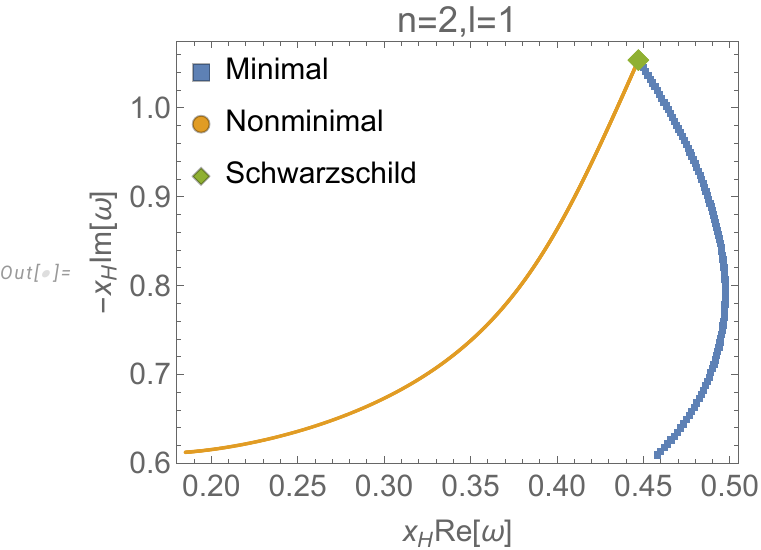}
        \caption{\empty}
        \label{fig:w21const}
    \end{subfigure}
    \begin{subfigure}{.32\textwidth}
        \centering
        \includegraphics[trim=1.5cm 0cm 0cm 0cm,clip=true,width=1\textwidth]{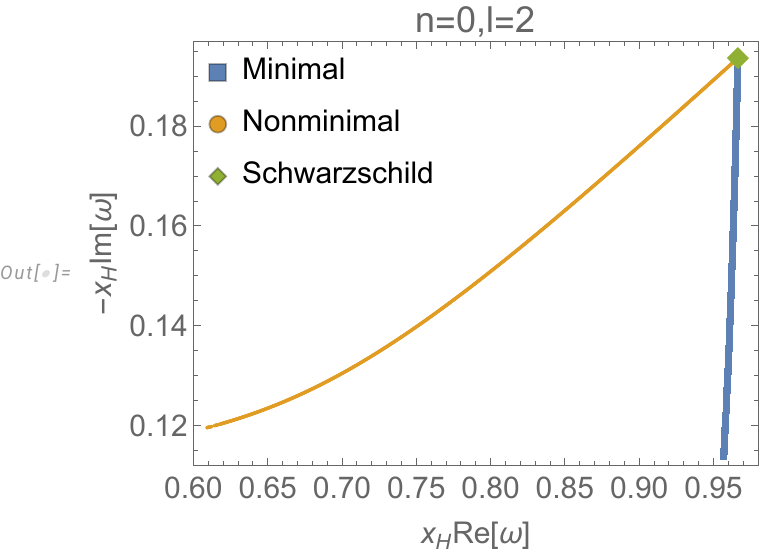}
        \caption{\empty}
        \label{fig:w02const}
    \end{subfigure}%
    \begin{subfigure}{.33\textwidth}
        \centering
        \includegraphics[trim=1.5cm 0cm 0cm 0cm,clip=true,width=1\textwidth]{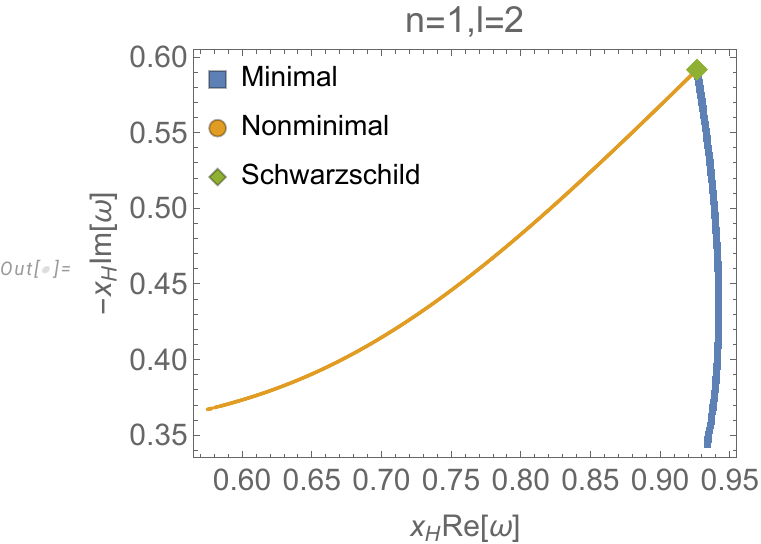}
        \caption{\empty}
        \label{fig:w12const}
    \end{subfigure}
    \begin{subfigure}{.3125\textwidth}
        \centering
        \includegraphics[trim=1.5cm 0cm 0cm 0cm,clip=true,width=1\textwidth]{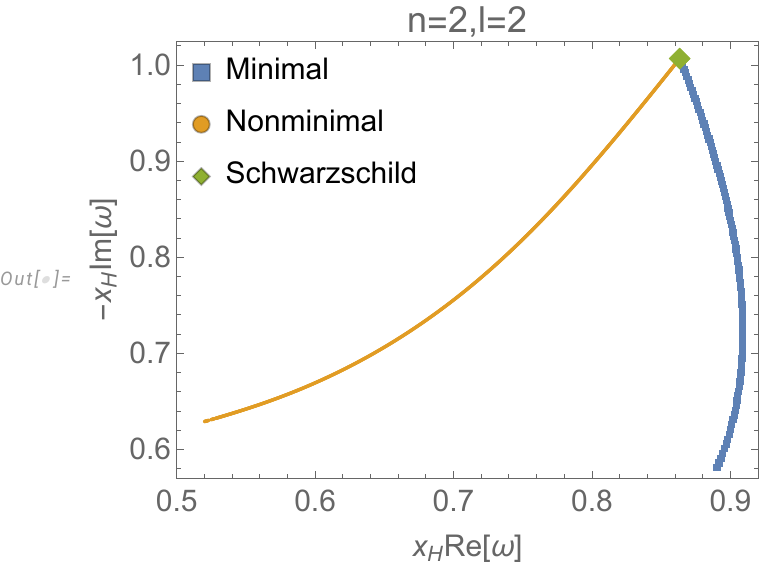}
        \caption{\empty}
        \label{fig:w22const}
    \end{subfigure}
    \caption{Positive branch of the eigenfrequencies with constant $\lambda$ for $l=0$ with (a) $n=0$, (b) $n=1$, and (c) $n=2$. The parameter $x_{\bar{\lambda}}/x_{\rm H}$ is varied continuously from $0$ to $0.99$ with the classical frequencies being $x_{\rm H} \omega_{00}^+=0.209294-i0.230394$, $x_{\rm H} \omega_{10}^+=0.178378-i0.70992$, $x_{\rm H} \omega_{20}^+=0.126957-i1.18915$, $x_{\rm H} \omega_{01}^+=0.582228-i0.196003$, $x_{\rm H} \omega_{11}^+=0.524424-i0.614865$, $x_{\rm H} \omega_{21}^+=0.447086-i1.05363$, $x_{\rm H} \omega_{02}^+=0.966422-i0.19361$, $x_{\rm H} \omega_{12}^+=0.926383-i0.59162$, and $x_{\rm H} \omega_{22}^+=0.863321-i1.00687$.
    The $\omega_{20}$ spectrum of nonminimal coupling (c) presents a gap in the imaginary part of the frequency.}
    \label{fig:wn0const}
\end{figure*}

\begin{figure}[!htb]
    \centering
    \includegraphics[trim=1.5cm 0cm 0cm 0cm,clip=true,width=0.9\columnwidth]{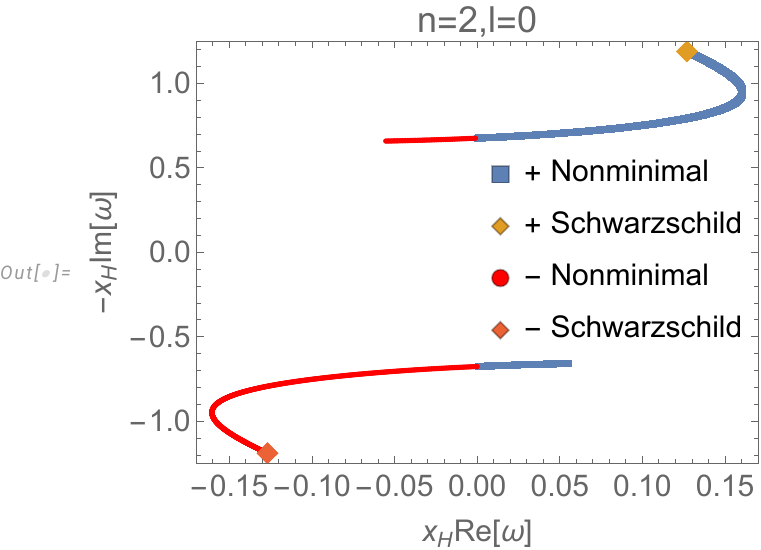}
    \caption{The two branches of eigenfrequencies $\omega^\pm_{20}$ with constant $\lambda$ for the mode $n=2$, $l=0$. The parameter $x_{\bar{\lambda}}/x_{\rm H}$ is varied continuously from $0$ to $0.99$. The upper half of the plot represents the observable eigenfrequency $\omega_{00}^{\rm (obs)}$.}
    \label{fig:w20const_twobranches}
\end{figure}

The most significant physical difference between the minimal and nonminimal couplings is in their frequency spectra which are compared in Fig.~\ref{fig:wn0const} for $l=0,1,2$ and $n=0,1,2$. The frequency spectra tend to have a more significant variation, especially of the real part, in nonminimal coupling compared to its minimal counterpart. Also, note the gap in the frequency spectrum for the mode $n=2$, $l=0$ of the nonminimally coupled case, Fig.~\ref{fig:w20const}.
The imaginary part of the frequency has a discontinuous jump from a typical negative value to a positive one, violating the boundary conditions (\ref{eq:Boundary cond - horizon}) and (\ref{eq:Boundary cond - Inf}).
The following sections will discuss the possible implications of such a phenomenon. Moreover, when considering the negative branch of the frequency eigenvalues $\omega^\pm$, the spectrum can be continuously extended to negative values of the real part of the frequency while preserving a negative imaginary part, which would now satisfy the boundary conditions and constitute the observable spectrum, see Fig~\ref{fig:w20const_twobranches}.

It is important to note that in all these plots the relevant parameter is varied continuously, but since
\begin{equation}
    \frac{x_{\bar{\lambda}}}{x_{\rm H}}= \frac{\bar{\lambda}^2}{1+\bar{\lambda}^2}\,,
\end{equation}
the true physical system requires only one point in each of the plots of frequency spectrum once the constant  $\bar{\lambda}$ is fixed.
In particular, since the classical limit is given by $\bar{\lambda}\to0$, only the first part of each plot is actually physically relevant since $\bar{\lambda}$ is expected to be small.
Note, for instance, that $x_{\bar{\lambda}}/x_{\rm H}=0.5$ already implies $\bar{\lambda}=1$, and $x_{\bar{\lambda}}/x_{\rm H}=0.9$ implies $\bar{\lambda}=3$.
The case $x_{\bar{\lambda}}/x_{\rm H}=1$ is not included in the graphs because it implies $\bar{\lambda}\to \infty$.

\subsection{Decreasing modification function}
\label{sec:QNM decreasing holonomy function}

In the decreasing case where $\lambda^2=\Delta/x^2$, $\chi=1$, the effective potential as well as the (anti)damping and modulation factors are given by

\begin{eqnarray}
    V_l (x) &=& 
    \left(1+\frac{\Delta}{x^2} \left(1-\frac{2 M}{x}\right)\right) \left(1-\frac{2M}{x}\right)
    \nonumber\\
    &&\qquad\times
    \left(\frac{l(l+1)}{x^2}
    + \frac{2M}{x^3}\right)
    \,,\\
    \zeta &=& \frac{\Delta}{x^3} \left(1+ \frac{\Delta}{x^2} \left(1-\frac{2 M}{x}\right)\right)^{-1/2}
    \nonumber\\
    &&\qquad\times
    \left(1-\frac{2 M}{x}\right) \left(1-\frac{3 M}{x}\right)
    \,,
\end{eqnarray}
\begin{eqnarray}
    \!\!\frac{\partial \zeta}{\partial x^*} \!\!&=&\!\! \zeta^2 - 3 \frac{\Delta}{x^4} \left(1-\frac{2M}{x}\right) \left(1-\frac{20M}{3x}+\frac{10M^2}{x^2}\right)
    \\
    U_l \!\!&=&\!\! \frac{1}{x^2} \left(1-\frac{2M}{x}\right) \Bigg[
    l(l+1)
    + \frac{2M}{x}
    \\
    \!\!&&\!\!
    + \frac{\Delta}{x^2}
    \left( 3+l(l+1)
    - \frac{2M}{x} l(l+1)
    + \frac{26M^2}{x^2}
    \right)
    \Bigg]
    \,,\nonumber\\
    Z \!\!&=&\!\! \sqrt{1+\frac{\Delta}{x^2}\left(1-\frac{2M}{x}\right)}
    \,.
\end{eqnarray}
Figs.~\ref{fig:NMpotmubar} and \ref{fig:NMdampmubar} contain the plot of the effective potential and the modulation factor, respectively, as a function of $x^*$ for  decreasing function of $\lambda$, i. e., $\lambda \propto 1/x$.

\begin{figure*}[!htb]
    \begin{subfigure}{.31\textwidth}
        \centering
        \includegraphics[trim=1.39cm 0cm 0cm 0cm,clip=true,width=1\textwidth]{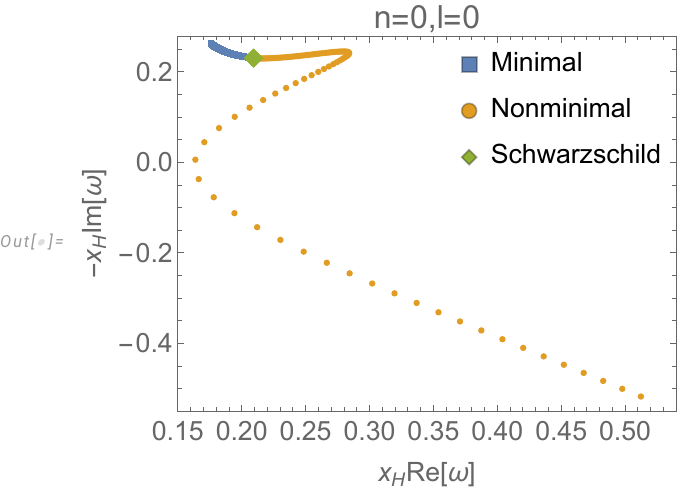}
        \caption{\empty}
        \label{fig:w00mubar}
    \end{subfigure}%
    \begin{subfigure}{.31\textwidth}
        \centering
        \includegraphics[trim=1.38cm 0cm 0cm 0cm,clip=true,width=1\textwidth]{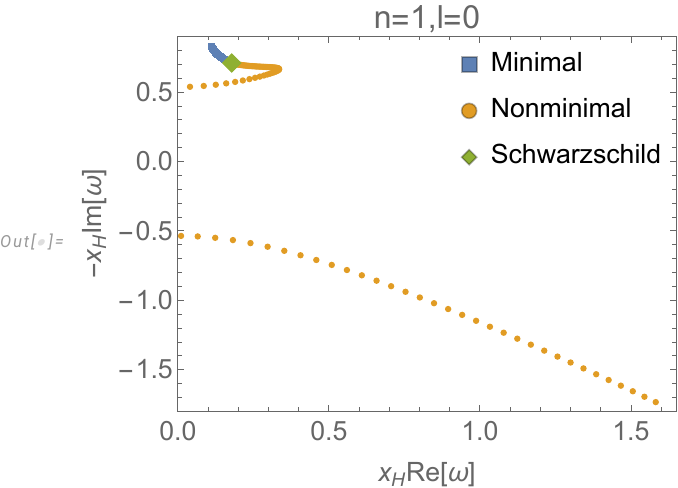}
        \caption{\empty}
        \label{fig:w10mubar}
    \end{subfigure}
    \begin{subfigure}{.3\textwidth}
        \centering
        \includegraphics[trim=1.38cm 0cm 0cm 0cm,clip=true,width=1\textwidth]{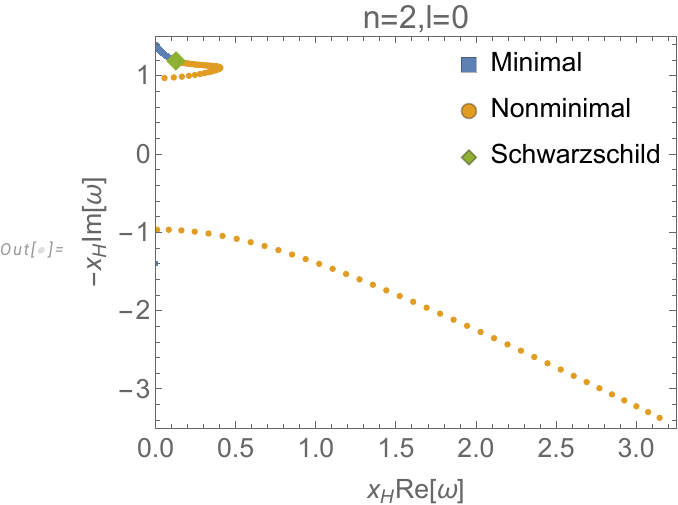}
        \caption{\empty}
        \label{fig:w20mubar}
    \end{subfigure}
    \begin{subfigure}{.32\textwidth}
        \centering
        \includegraphics[trim=1.4cm 0cm 0.28cm 0cm,clip=true,width=1\columnwidth]{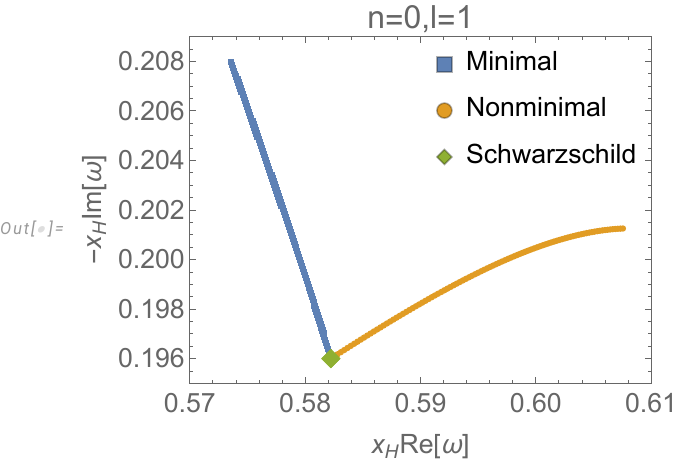}
        \caption{\empty}
        \label{fig:w01mubar}
    \end{subfigure}
    \begin{subfigure}{.305\textwidth}
        \centering
        \includegraphics[trim=1.4cm 0cm 0cm 0cm,clip=true,width=1\columnwidth]{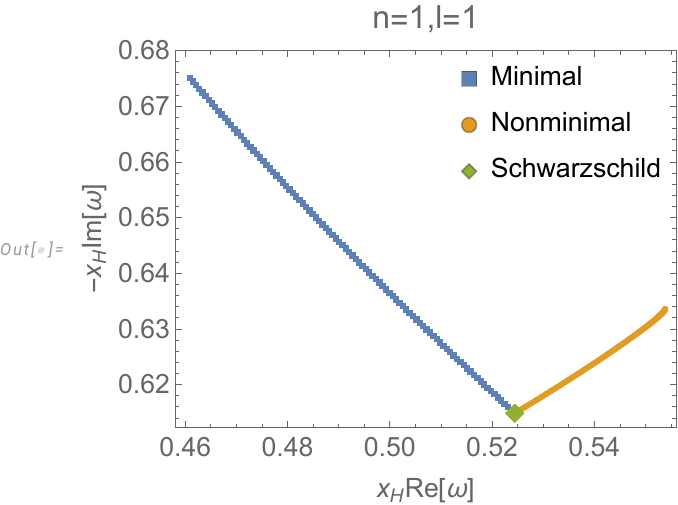}
        \caption{\empty}
        \label{fig:w11mubar}
    \end{subfigure}
    \begin{subfigure}{.32\textwidth}
        \centering
        \includegraphics[trim=1.4cm 0cm 0cm 0cm,clip=true,width=1\columnwidth]{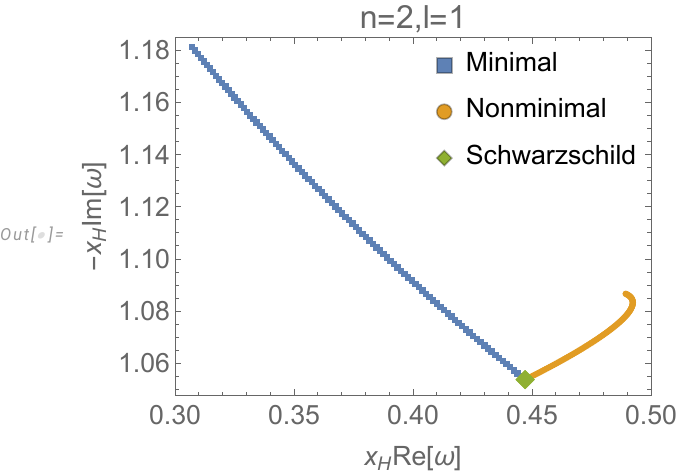}
        \caption{\empty}
        \label{fig:w21mubar}
    \end{subfigure}
    \begin{subfigure}{.32\textwidth}
        \centering
        \includegraphics[trim=1.4cm 0cm 0cm 0cm,clip=true,width=1\textwidth]{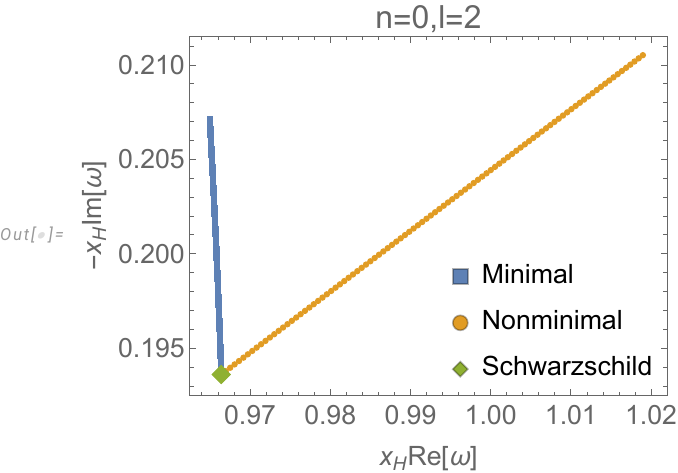}
        \caption{\empty}
        \label{fig:w02mubar}
    \end{subfigure}%
    \begin{subfigure}{.31\textwidth}
        \centering
        \includegraphics[trim=1.4cm 0cm 0cm 0cm,clip=true,width=1\textwidth]{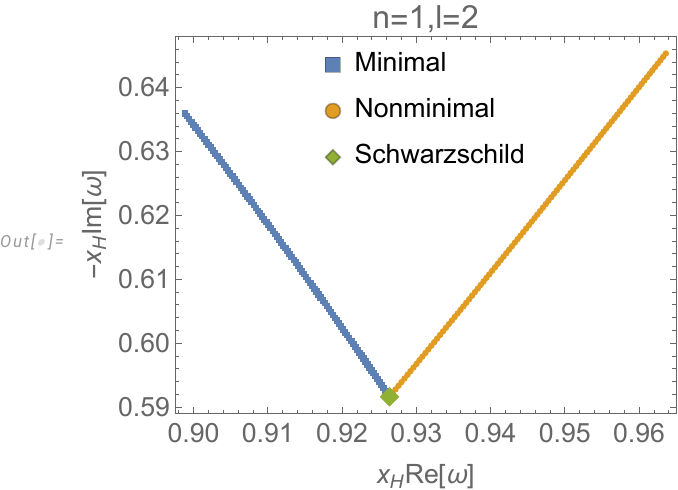}
        \caption{\empty}
        \label{fig:w12mubar}
    \end{subfigure}
    \begin{subfigure}{.31\textwidth}
        \centering
        \includegraphics[trim=1.4cm 0cm 0cm 0cm,clip=true,width=1\textwidth]{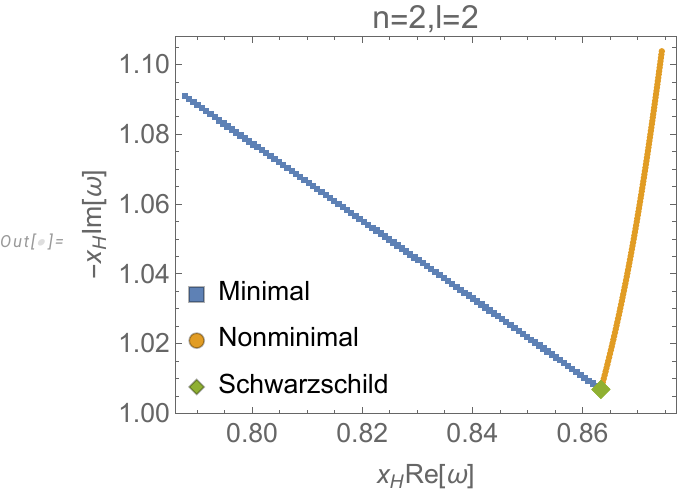}
        \caption{\empty}
        \label{fig:w22mubar}
    \end{subfigure}
    \caption{Positive branch of the eigenfrequencies with decreasing $\lambda\propto 1/x$. The parameter $\Delta/x_{\rm H}^2$ is varied from $0$ to $1$. The imaginary part of the $n=l=0$ mode in (a) vanishes at $\Delta/x_{\rm H}^2\approx 0.77142$.}
    \label{fig:wn0mubar}
\end{figure*}

The frequency spectra for nonminimal and minimal coupling are compared in Fig.~\ref{fig:wn0mubar} for $l=0,1,2$ and $n=0,1,2$.
These spectra differ significantly from the constant $\lambda$ case.

Contrary to the constant case, the entirety of the frequency spectrum plots is relevant because the parameter
\begin{eqnarray}
    \frac{\Delta}{x_{\rm H}^2} = \frac{\Delta}{4M^2}\,,
\end{eqnarray}
is mass dependent for constant $\Delta$.
This means that the QNM spectra will differ significantly, for instance, between different stages of a black hole evaporation process and could be used to track it.
In particular, assuming $\Delta$ is of the order of the Planck area, the last part of the plots are relevant once the black hole has reached Planckian mass.

\begin{figure*}[ht]
    \begin{subfigure}{.32\textwidth}
        \centering
        \includegraphics[trim=1.4cm 0cm 0cm 0cm,clip=true,width=1\textwidth]{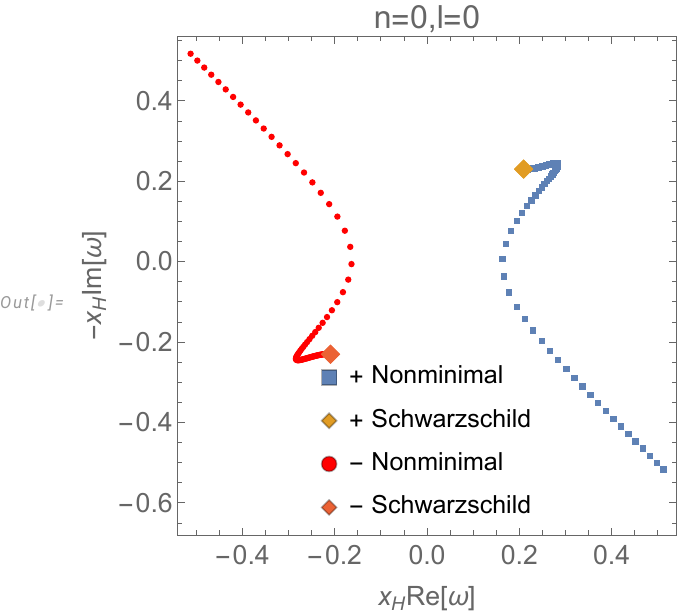}
        \caption{\empty}
        \label{fig:w00mubar_twobranches}
    \end{subfigure}
    \begin{subfigure}{.32\textwidth}
        \centering
        \includegraphics[trim=1.4cm 0cm 0cm 0cm,clip=true,width=1\textwidth]{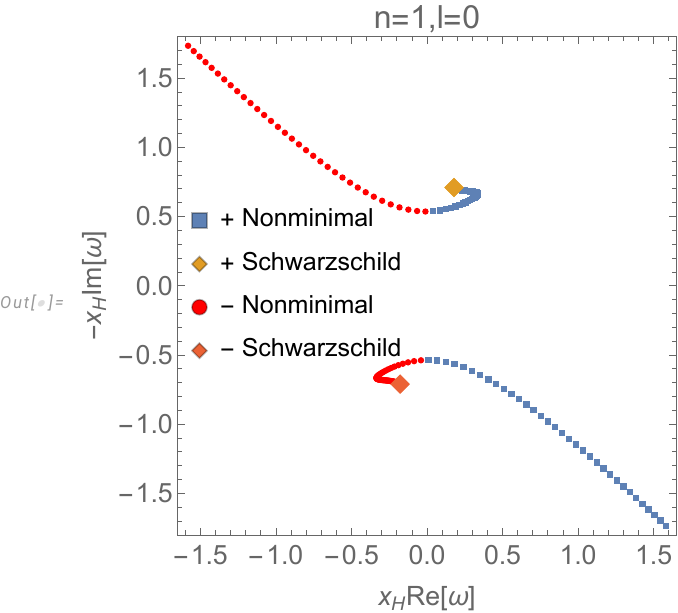}
        \caption{\empty}
        \label{fig:w10mubar_twobranches}
    \end{subfigure}
    \begin{subfigure}{.32\textwidth}
        \centering
        \includegraphics[trim=1.5cm 0cm 0cm 0cm,clip=true,width=1\textwidth]{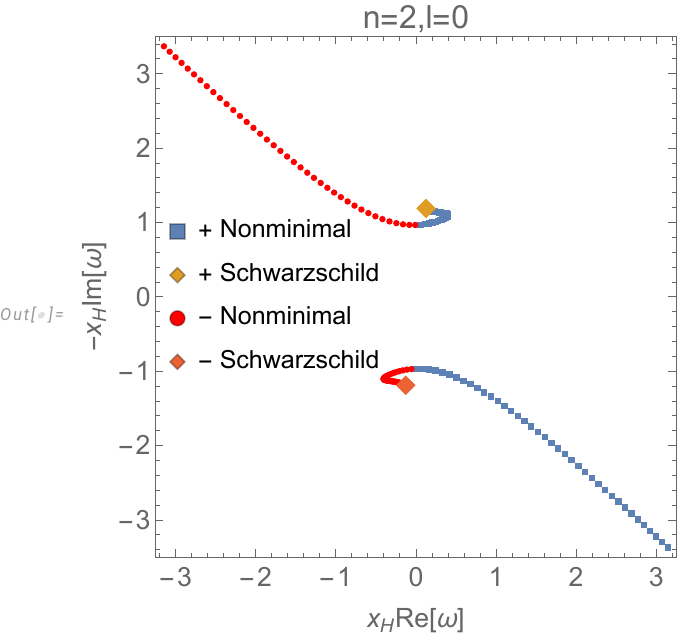}
        \caption{\empty}
        \label{fig:w20mubar_twobranches}
    \end{subfigure}
    \caption{The two branches of the eigenfrequencies $\omega^\pm_{n0}$ with decreasing $\lambda\propto 1/x$ for $l=0$ with (a) $n=0$ (b) $n=1$, and (c) $n=2$. The parameter $\Delta/x_{\rm H}^2$ is varied from $0$ to $1$. The upper half of the plots in (b) and (c) represent the respective observable eigenfrequencies $\omega_{10}^{\rm (obs)}$ and $\omega_{20}^{\rm (obs)}$. In contrast, while the negative branch $\omega_{00}^-$ in (a) acquires negative imaginary component, it remains disconnected from its positive branch $\omega_{00}^+$.}
    \label{fig:wn0mubar_twobranches}
\end{figure*}

The imaginary part of the frequencies with $l=0$ and $n=1$ or $n=2$ change sign for sufficiently small black hole mass, see Fig.~\ref{fig:wn0mubar}.
The positive branch no longer satisfies the boundary conditions (\ref{eq:Boundary cond - horizon}) and (\ref{eq:Boundary cond - Inf}) in the spectrum region with ${\rm Im[\omega^+]}>0$.
As explained in the previous Subsection, the spectrum can be continuously extended to negative values of the real part of the frequency while preserving a negative imaginary part, and hence still satisfying the boundary conditions, by gluing it to the negative branch of the frequency eigenvalues, see Figs~\ref{fig:w10mubar_twobranches} and \ref{fig:w20mubar_twobranches}.

While the spectrum with the positive imaginary part of the frequencies violates the boundary conditions necessary for observable QNMs in the asymptotic region, it may still have important near-horizon physics. Upon the substitution in $e^{-i\omega^\pm (t-x^*)}$ for outgoing solutions, we find that the modes with ${\rm Im}[\omega^\pm]>0$ do not decay exponentially with time but are instead amplified. In contrast, they decay with the distance scale $x^*$, suggesting that such modes may have relevant effects only near the horizon. In particular, note that the positive branch of the $\omega_{00}$ mode, shown in Fig.~\ref{fig:w00mubar}, has an imaginary part that continuously goes through zero while keeping a finite and positive real part, indicating that this could be an actual destabilizing mode.
While its negative branch frequency $\omega_{00}^-$ acquires negative imaginary component, it remains disconnected from its positive branch $\omega_{00}^+$ and is not possible to implement an extension similar to that of $\omega_{10}^{\rm (obs)}$ and $\omega_{20}^{\rm (obs)}$, see Fig.~\ref{fig:w00mubar_twobranches}.
We elaborate on our speculations of the physical implications of these observations in Section~\ref{sec:Discussion}.

\subsection{High frequency modes}
\label{sec:High frequency modes}

The WKB approximation works well for highly oscillatory solutions, such that $\omega x^*$ is large. Therefore, the approximation can be used for small $x^*$ provided $\omega$ is high. Since the small-$x$ behavior is markedly different between GR and EMG --- the black-hole singularity is replaced by a minimum-radius hypersurface in EMG --- one would expect that the high-frequency modes may differ for these two cases.
In the following, we make use of the expansions near the minimum-radius surface derived in Appendix~\ref{app:Near minimum radius surface}.

\subsubsection{Nonminimal coupling}
Near the minimum-radius hypersurface, $x=x_\lambda+r=$ with $r/x_\lambda\ll1$, the nonminimally coupled potential (\ref{eq:Potential - Schrodinger}) becomes, to leading order in $r/x_\lambda$,
\begin{eqnarray}\label{eq:General potential - near minimum radius}
    U_l =
    - \sigma + \rho \frac{r}{x_\lambda}
    =: - \sigma + \rho^* (x^*)^2
\end{eqnarray}
where $\sigma$ and $\rho$ are complicated expressions of $\lambda$ and may depend on $l$. Their detailed expressions will not be necessary for the following, except that they are finite and positive.
In particular, there is no divergence as $r\to0$, provided $x_\lambda$ is finite.
For the constant case $\lambda(x)=\bar{\lambda}$ with $\chi=1/\sqrt{1+\bar{\lambda}^2}$ we obtain
\begin{eqnarray}
    \!\!\!\!\sigma_{\bar{\lambda}} \!\!&=&\!\!
    \frac{6 M^2}{x_{\bar{\lambda}}^4}
    - \frac{5 M}{x_{\bar{\lambda}}^3}
    + \frac{1}{x_{\bar{\lambda}}^2}
    \,,\\
    \!\!\!\!\rho_{\bar{\lambda}} \!\!&=&\!\!
    \frac{26 M^2}{x_{\bar{\lambda}}^4}
    - \frac{2 \left(l(l+1)+9\right) M}{x_{\bar{\lambda}}^3}
    + \frac{l(l+1)+3}{x_{\bar{\lambda}}^2}
    \,.
\end{eqnarray}
and for the decreasing case $\lambda(x)=\sqrt{\Delta/x^2}$ with $\chi=1$ we obtain
\begin{eqnarray}
    \sigma_\Delta \!\!&=&\!\!
    \frac{30 M^2}{x_\Delta^4}
    - \frac{20 M}{x_\Delta^3}
    + \frac{3}{x_\Delta^2}
    \,,\\
    \rho_\Delta \!\!&=&\!\! \bigg[\frac{396 M^3}{x_\Delta^4}
    - \frac{4 \left(3 l^2+3 l+100\right) M^2}{x_\Delta^3}
    - \frac{2 \left(l(l+1)+6\right)}{x_\Delta}
    \nonumber\\
    \!\!&&\!\!\qquad
    + \frac{2M \left(5 l^2+5 l+63\right)}{x_\Delta^2}
     \bigg] 
    \frac{1}{\left(2 M-x_\Delta\right)} \nonumber
    \,.
\end{eqnarray}
In contrast, the classical potential (\ref{eq:Classical potential}) near the singularity is given by
\begin{equation}\label{eq:Classical potential near singularity}
    V_l = - \frac{4 M^2}{r^4}\,,
\end{equation}
in an expansion to the leading order of $x=r$. Inserting the potential (\ref{eq:General potential - near minimum radius}) into (\ref{eq:QNM equation - Schrodinger - WKB}), and using (\ref{eq:Tortoise minimum radius}), defining
$$z=x^*-c^*,$$
we obtain the equation
\begin{equation}
    \frac{\partial^2 S_{lm}}{\partial z^2} + \left[\omega^2 + \sigma - \frac{\gamma \rho}{4 x_\lambda^2 \lambda_\lambda^4} z^2\right] S_{lm} = 0
    \,,
\end{equation}
or
\begin{equation}\label{eq:QNM equation - Schrodinger - WKB - near minimum radius}
    \frac{\gamma}{x_\lambda} \frac{1}{\lambda_\lambda^4} \left( r \frac{\partial^2 S_{lm}}{\partial r^2} + \frac{1}{2} \frac{\partial S_{lm}}{\partial r} \right) + \left[\omega^2+\sigma - \frac{\rho}{x_\lambda} r\right] S_{lm} = 0
    \,.
\end{equation}
Equation (\ref{eq:QNM equation - Schrodinger - WKB - near minimum radius}) has a regular singular point at $r=0$. Defining
\begin{equation}\label{eq:Near minimum radius solution}
    S_{lm}= e^{- r \lambda_\lambda^2 \sqrt{\rho/\gamma}} \sqrt{r}\; \tilde{S}_{lm}
\end{equation}
equation (\ref{eq:QNM equation - Schrodinger - WKB - near minimum radius}) becomes
\begin{equation}\label{eq:QNM equation - Schrodinger - WKB - near minimum radius - simp}
    w \frac{\partial^2 \tilde{S}_{lm}}{\partial w^2}
    + \left[b - w\right] \frac{\partial \tilde{S}_{lm}}{\partial w}
    - a \tilde{S}_{lm} = 0
    \,,
\end{equation}
where
\begin{eqnarray}
    w &=& 2\lambda_\lambda^2 \sqrt{\frac{\rho}{\gamma}} r
    \,,\\
    b &=& \frac{3}{2}
    \,,\\
    a &=& \frac{3}{4}-\frac{x_\lambda \lambda_\lambda^2}{2\sqrt{\gamma\rho}}(\omega^2+\sigma)\,.
\end{eqnarray}
Equation (\ref{eq:QNM equation - Schrodinger - WKB - near minimum radius - simp}) is Kummer's equation, which has the general solution~\cite{NISThandbook,Kummereqs}
\begin{eqnarray}\label{eq:Kummer solution}
    \tilde{S}_{lm} = \frac{c_M M (a,b,w)}{(-a)^{(1-b)/2} \Gamma(b)}
    + \frac{c_U U (a,b,w)}{\Gamma \left((b+1)/2-a\right)}
    \,,
\end{eqnarray}
where $c_M$ and $c_M$ are constants, $M$ is the Kummer function and $U$ is the Tricomi confluent hypergeometric function.

For large $a>0$ with $b$ and $w$ fixed, these functions approximate\footnote{Readers may refer to Eqs. 13.5.13 and 13.5.15 in Ref.~\cite{Kummereqs}}:
\begin{widetext}
\begin{eqnarray}
    M(a,b,w) &\approx& \Gamma(b) e^{w/2} \left((b/2-a)w\right)^{(1-b)/2} J_{b-1} \left(\sqrt{(2b-4a)w}\right)
    \,,\\
    U(a,b,w) &\approx& \Gamma\left((b+1)/2-a\right) e^{w/2} w^{(1-b)/2}
    \left[ \cos(a\pi) J_{b-1} \left(\sqrt{(2b-4a)w}\right)
    - \sin (a\pi) Y_{b-1} \left(\sqrt{(2b-4a)w}\right)\right]
    \nonumber\\
    &=& \Gamma\left((b+1)/2-a\right) e^{w/2} w^{(1-b)/2} J_{1-b} \left(\sqrt{(2b-4a)w}\right)
    \,,
\end{eqnarray}
\end{widetext}
where $J_\alpha(y)$ and $Y_\alpha(y)$ are the Bessel functions of the first and second kind, respectively, and the last equality holds for noninteger $b$, as is our case $b=3/2$.
In turn, the asymptotic form of the Bessel function for large $y$ is given by
\begin{eqnarray}
    J_\alpha(y) &\approx& \sqrt{\frac{2}{\pi y}} \cos \left(y-\frac{\pi}{2}\left(\alpha+\frac{1}{2}\right)\right)
    \,,
    \,,
\end{eqnarray}

At large frequencies we have
\begin{equation}\label{eq:High freq a}
    a \approx - \frac{x_\lambda \lambda_\lambda^2}{2\sqrt{\gamma\rho}} \omega^2
\end{equation}
large and positive provided $\omega^2<0$. Then,
$a\gg b=3/2$, and, using (\ref{eq:Tortoise minimum radius}), the argument of the Bessel functions can be approximated as
\begin{equation}
    \sqrt{- 4a w} \approx - \omega z\,.
\end{equation}
Using (\ref{eq:High freq a}), the high frequency assumption $a\gg1$ means
\begin{eqnarray}
   - \omega^2 \gg  \frac{2}{x_\lambda \lambda_\lambda^2} \sqrt{\gamma\rho}
\end{eqnarray}
which for the constant and decreasing $\lambda$ cases respectively translates to
\begin{eqnarray}
    -\omega^2_{\bar{\lambda}} &\gg&  \frac{\sqrt{26}}{x_{\rm H}^2} \left(\frac{x_{\rm H}}{x_{\bar{\lambda}}}\right)^4
    \,,\\
   - \omega^2_\Delta &\gg&  \frac{6}{x_{\rm H}^2} \sqrt{\frac{33}{2}}\; \left(\frac{x_{\rm H}}{x_\Delta}\right)^4
\end{eqnarray}
where we have assumed $2M/x_\lambda\gg1$.
This implies that $|{\rm arg}[\omega]|\approx\pi/2$ or ${\rm Im}[\omega]\gg{\rm Re}[\omega]$, and therefore belong to very short-lived modes.

Using all the above we find that the solution (\ref{eq:Near minimum radius solution}) at high frequencies and in the asymptotic region becomes
\begin{equation}
    S_{lm} \sim \left[c_M e^{-i\pi/2}
    + c_U \right] e^{- i\omega z}
    + \left[c_M e^{+i\pi/2}
    + c_U\right] e^{i\omega z}
    \,.
\end{equation}
Avoiding outgoing waves at the boundaries implies the condition \cite{Nietzke}
\begin{equation}\label{eq:High freq bound cond}
    c_U = - c_M e^{-i\pi/2}
    \,.
\end{equation}

We extend the coordinate $z=x^*-c^*$ to the complex plane.
Also extending $r$ into the complex plane and writing $r=|r|e^{i{\rm arg}[r]}$, from (\ref{eq:Tortoise minimum radius}) we find that near the minimum-radius hypersurface we have ${\rm Re}[z]>0$ in the region $\pi<{\rm arg}[r]<3\pi$.
A rotation by $2\pi$ in the $r$ plane becomes a rotation by $\pi$ in the $z$ plane, in contrast to GR which has a $3\pi/2$ rotation in $r$ and a $3\pi$ rotation in the $z$ plane since $z\approx -r^2/2$.
Another significant difference from GR is that, unlike $z\sim r^2$, the square root $z\sim r^{1/2}$ is a multivalued function and hence a monodromy of $r^{1/2}$ around $r=0$ causes a $2\pi i$ discontinuity in the path in the $z$-plane upon crossing the branch cut placed between the regular singular points $x=x_\lambda$ and $x=2M$, which leads us to take the phase $e^{i3\pi} z$ rather than just $e^{i\pi}z$.

To compute the monodromy, we will use the property\footnote{Note that the extra contributions of monodromies of the Kummer functions would be negligible for high frequencies. Hence, they are not considered.}:
\begin{eqnarray}
    J_\alpha(z e^{i\pi m}) = e^{i\pi m \alpha} J_\alpha(z) \, .
\end{eqnarray}
Therefore, we get:
\begin{eqnarray}
    S_{lm} &\sim& \left[c_M e^{i5\pi/2} + c_U \right] e^{- i\omega z}
    + \left[c_M e^{i7\pi/2} + c_U \right]e^{+ i \omega z}
    \nonumber\\
    &=& 2 c_M e^{i\pi/2} e^{- i\omega z}
    \,,
\end{eqnarray}
where we used (\ref{eq:High freq bound cond}) in the second line.
The monodromy around $r=0$ is therefore unity
\begin{equation}
    \frac{c_M e^{i5\pi/2} + c_U}{c_M e^{i\pi/2}
    + c_U} = 1\,,
\end{equation}
which is identical to the classical result \cite{Nietzke}, using $j=1$ in that reference.

On the other hand, the tortoise coordinate near the horizon (\ref{eq:Tortoise horizon}) is nearly classical, as is the mode function (\ref{eq:Boundary cond - horizon}), which we rewrite here as
\begin{equation}
    S_{lm} \sim e^{-i\omega z}
    \approx \exp \left(\frac{-2i\omega M}{\chi} \ln \left(x-2M\right)\right)
    \,,
\end{equation}
The monodromy of a contour around the horizon, $x=2M$, must take into account a $2\pi i$ discontinuity from the logarithm being multivalued, leading to take the phase $e^{-i4\pi}$ and therefore obtain the monodromy $\exp\left(8\pi\omega M/\chi\right)$.
Equating the two monodromies we obtain
\begin{equation}
    e^{8\pi\omega M/\chi} = 1
    \,,
\end{equation}
and hence
\begin{equation}\label{eq:High frequency evalue}
    \omega_n = - i \frac{\chi \, n}{4 M} \,,
\end{equation}
for some (positive, large) integer $n$, which is identical, up to the appearance of $\chi$, to the classical result \cite{Nietzke}, using $j=1$ in that reference.
This result is independent of the  parameter $\lambda$, except for the appearance of the global factor $\chi$ which is unity for asymptotically vanishing $\lambda$.
Furthermore, the lack of a real part in the frequency indicates that these high frequency modes describe purely evanescent solutions rather than oscillatory.
In GR this seems to be a feature of scalar QNMs but fields of different spin, for instance electromagnetic QNMs, may exhibit high frequency modes with a nonvanishing real part for spacetimes with generic singularities \cite{Das_2005,Nietzke}.

\subsubsection{Minimal coupling}

Near the minimum radius surface, $x=x_\lambda+r=$ with $r/x_\lambda\ll1$, the minimally coupled potential (\ref{eq:Minimal pot}) acquires, to leading order in $r/x_\lambda$, the same qualitative form as the nonminimally coupled one,
\begin{eqnarray}
    V_l &\approx& \sigma_{\rm min} + \rho_{\rm min} \frac{r}{x_\lambda}\,,
\end{eqnarray}
and therefore the same results apply.

\section{Discussion}
\label{sec:Discussion}

We have presented the first detailed analysis of scalar quasinormal modes in emergent modified gravity, considering different types of correction functions and nonminimal coupling. These models are the most general modifications of a spherically symmetric background subject to second-order field equations, avoiding common instabilities related to higher derivative terms. Because they rely on details of the canonical formulation of gravity, they can also be used as effective theories of canonical quantum gravity or, specifically, loop quantum gravity. In the latter case, the key modification function included in our analysis is interpreted as the length parameter of holonomies in a covariant lattice formulation of gravity and thus signals implications of a specific type of discrete geometries.

We found several new characteristic features compared with previous studies of this general problem. The main ones include a new effect in the presence of nonminimal coupling of the scalar field that may enhance the amplitude of some modes, making them more likely to be detected or to be restricted by observational upper bounds. This enhancement in amplitude due to nonminimal coupling is a significant finding as it can potentially lead to the detection of modes that were previously undetectable. Nonminimal coupling of a scalar field is often considered unnatural, but it is an essential contribution if the scalar is used as a model for the more complicated case of gravitational quasinormal modes. In the latter case, only one space-time geometry is implied by the combination of a spherically symmetric background and a superposition of quasinormal modes, which have been separated only in the formal process of solving equations. Nonminimal coupling is then merely a way of including generic modifications of the classical gravitational interaction.
Therefore, it is interesting to see that this level of generality leads to new physical effects. 
In addition, the frequency spectra of quasinormal modes change in characteristic ways for the models considered here. Compared with previous studies of quasinormal modes in modified gravity, including loop quantum gravity models, such as \cite{QNMModified}, our results show a more significant variation of the real parts of frequencies.

We found that scalar quasinormal modes can be set up as perturbations on a gravitational background in EMG, following the procedure familiar from the GR treatment.
We used the WKB approximation to compute the QNMs spectra for two classes of models of scalar matter coupled to EMG, given by minimal and nonminimal coupling to the space-time background determined by the emergent metric.
Additionally, each coupling depends on a modification function $\lambda$ whose dependence on the areal radius is undetermined by theory except for the property that the limit $\lambda\to0$ recovers GR. (Specific approaches to quantum gravity may be used to motivate a specific class of modification functions in an effective theory.)
We thus computed the QNMs spectra under two common choices of $\lambda$ --- a constant and a decreasing function that vanishes asymptotically.
The spectra of the two couplings and the two choices of $\lambda$ differ significantly, making it a robust test against observational data for each model.
Additionally, more sophisticated numerical methods will improve the results obtained by the WKB approximation, allowing better discrimination between models.
While currently there is much freedom in the choice of $\lambda$ to fit the QNM spectrum, the results can be used for consistency between multiple observations to fix a preferred $\lambda$ function ultimately.

We have found that nonminimal coupling allows vanishing real and imaginary frequency components and also presents a richer spectrum of frequencies with the opposite sign of their imaginary part in the case of a decreasing function $\lambda(x)=\sqrt{\Delta/x^2}$ where the eigenfrequencies are mass dependent.
At the masses where ${\rm Re}[\omega]=0$, the modes become evanescent as they decay but do not propagate, a possible consequence of which is that no QNMs would be observable at such stage of the evaporation process.
On the other hand, a possible consequence of the solutions with a positive imaginary part of the frequency is superradiant instability, accelerating the evaporation process since the energy to amplify the waves can only come from the black hole's mass, potentially leading to a phenomenon akin to black hole bombs described in \cite{BHbomb}. This is not a unique phenomenon of EMG given that \cite{BHbomb} studies this idea in the context of GR, but the modifications to the spectrum in EMG would certainly change the dynamics of the black hole bombs.
In particular, note that the positive branch of the $\omega_{00}$ mode, shown in Fig.~\ref{fig:w00mubar}, continuously goes through zero, changing the sign of its imaginary part. In contrast, its real part remains finite and positive at the parameter value $\Delta/x_{\rm H}^2\approx 0.77142$.
This indicates that this mode is potentially destabilizing, providing a mechanism for the black hole bomb once its mass has evaporated to the critical value $M\approx 0.56928\sqrt{\Delta}$.
Also note that for masses near such critical value, the $\omega_{00}$ modes are long-lived and still propagate due to ${\rm Re}[\omega_{00}]>0$, providing a characteristic signature of the imminent start of the explosion.
This black hole bomb hypothesis could also imply a reduced prevalence of primordial black holes, which would have potential cosmological implications.
However, the perturbative treatment is not enough to model the full dynamics of such amplified waves; a nonperturbative treatment will be needed to describe them consistently.

The potential of more intricate versions of scalar matter Hamiltonians in EMG has numerous interesting implications. For instance, it is possible to add modifications of the scalar field where $\phi$ appears as an argument of trigonometric functions in the Hamiltonian constraint, such as $\sin (\nu \phi)$ (sometimes motivated by loop quantum gravity and referred to as point holonomy modifications). In this case,  $\nu$, like $\lambda$, may be scale dependent.
Understanding the potential changes in the QNMs spectra in the presence of these modifications is one possible direction to pursue.
As pointed out in Subsection~\ref{sec:Nonminimal coupling}, the global factor $\chi$ is allowed to depend on the scalar field $\phi$. This dependence is crucial for introducing the point holonomy modifications in a covariant way, as discussed in \cite{EMGscalar}.
It can result in nontrivial modifications to the high-frequency eigenvalues given that $\chi(\phi)$ appears in the expression (\ref{eq:High frequency evalue}) and therefore high frequencies could be sensitive to the point holonomy parameter $\nu(x)$.
However, the point holonomy modifications are relevant only beyond leading order in the test field and, hence, would have minor effects on the perturbative treatment.
On the other hand, these modifications can significantly contribute to a nonperturbative treatment of the unstable modes, potentially altering the dynamics of the hypothetical black hole bombs, a prospect that is both exciting and promising.

Furthermore, electromagnetic coupling to emergent modified gravity is possible where a black hole solution with electric charge exists \cite{alonsobardaji2023Charged,EmEM}.
A robust nonsingular behavior of the black hole requires modified gravity and electromagnetism~\cite{Johnson:2023skw}, which is possible only by introducing the concept of the emergent electric field \cite{EmEM}.
Similar to how EMG distinguishes gravity from emergent spacetime, here a distinction is made between the canonical electric momentum and the emergent electric field. The latter enters the observations via the Lorentz force as a component of the strength tensor. An analysis of QNMs of charged black holes could help study the emergent electric field hypothesis.

Future work will be dedicated to gravitational QNMs.
These are more complicated to deal with than those of scalar matter.
In particular, a perturbative treatment around the spherically symmetric background must first be developed for EMG, a most difficult task since, to our knowledge, not even the classical treatment exists in the canonical form to this day.
Given that in GR, the Regge-Wheeler potential of a spin $j$ field is given by \cite{ReggeWheeler}
\begin{equation}
    V_{lj} (x) = \frac{1}{x^2} \left(1-\frac{2M}{x}\right) \left(l(l+1)+\frac{1-j^2}{x}\right)\,,
\end{equation}
a first step to get qualitative results for \emph{axial} gravitational QNMs is to modify our potentials in this work to incorporate the spin $j=2$.
This was done for instance in \cite{Gingrich_2024} to compute the QNMs for all massless spin perturbations, including gravitational ones and spin-1/2, for the minimal coupling with constant $\lambda$.
The results of this procedure would remain qualitative only and seen with great skepticism because it is not guaranteed that they are gravitational perturbations, or perturbations of different spin fields, in EMG, nor that they preserve the covariance of the theory if considered gravitational perturbations, which require covariance conditions different from those of the scalar field.

It is tempting to devise a simple but rigorous minimal coupling for the gravitational waves as follows.
Consider the emergent metric $\bar{\tilde{g}}$ as the background and include perturbations $h_{\mu \nu}$ such that the full metric is given by
\begin{equation}\label{eq:Metric with perturbations}
    \tilde{g}_{\mu \nu} = \bar{\tilde{g}}_{\mu\nu} + h_{\mu \nu}
    \,,
\end{equation}
where the components of the perturbation may be written as
\begin{eqnarray}
    h_{\mu \nu} {\rm d} x^\mu {\rm d} x^\nu &=& - {\cal N}^2 {\rm d} t^2
    \\
    &&
    + \beta^{-1} h_{ab} \left({\rm d} x^a + {\cal N}^a{\rm d}t\right)\left({\rm d} x^b + {\cal N}^b{\rm d}t\right)
    \,,\nonumber
\end{eqnarray}
and $\beta$ is the same function used in (\ref{eq:Spacetime metric - modified - Schwarzschild})--- though, it is not necessary to include it as it could be absorbed by $h_{ab}$.
We now define the minimal coupling of the perturbations with the emergent background by the effective Einstein equations
\begin{equation}\label{eq:Effective Einstein}
    G_{\mu\nu} [\tilde{g}] = G_{\mu\nu} [\bar{\tilde{g}}]
    \,.
\end{equation}
Performing an expansion in $h_{\mu\nu}$ we obtain
\begin{equation}
    G_{\mu\nu} [\tilde{g}] = G_{\mu\nu}[\bar{\tilde{g}}]
    + G^{(h)}_{\mu\nu}[h]
    + O (h^2)
    \,,
\end{equation}
and plugging into (\ref{eq:Effective Einstein}), we obtain the equations of motion for the perturbations to the leading order
\begin{equation}\label{eq:Effective Einstein pert}
    G^{(h)}_{\mu\nu}[h] = 0
    \,.
\end{equation}
One may then choose the usual Regge-Wheeler gauge to obtain the equations of motion of the $h$.
This, however, results in three independent equations of motion for two degrees of freedom, unlike in GR where only two equations are obtained and is, therefore, an over-constrained system that has no fully consistent solutions unless $\lambda\to0$, which is simply the GR limit.
Therefore, the full canonical treatment of perturbations within the EMG formulation is needed to obtain consistent modified equations of motion for the perturbative degrees of freedom.
This seems like a natural conclusion given that the gravitational background itself does not follow Einstein's equations; therefore, the perturbations should not be expected to follow them either.
While a systematic treatment of gravitational perturbations in canonical form exists \cite{PertLQG}\textemdash\, which has been applied extensively to simpler cosmological models with scalar perturbations but has not been adapted to spherical symmetry with tensor perturbations\textemdash\, experience shows that it will take large amounts of effort and time before we get the perturbative EMG Hamiltonian.
However, the results presented here for scalar matter QNMs are encouraging, as they pave the way for the expectation that the gravitational QNMs spectrum in EMG would also present significant deviations from GR. This opens up the exciting possibility of testing EMG in the strong gravity regimes in the near future.

\section*{Acknowledgements}
The work is supported by SPARC MoE grant SPARC/2019-2020/P2926/SL and by NSF grant PHY-2206591. The international travel of EID was supported by IITB-IOE-SCPP Grant (IOE00I0-145). MB and EID thank IITB for hospitality during the early stages of this project.

\begin{appendix}
\section{Spherically symmetric emergent modified gravity}
\label{app:Spherical emergent modified gravity}

\subsection{Vacuum}
\label{app:Vacuum}

\subsubsection{Canonical system}

According to EMG, the most general, covariant Hamiltonian constraint in a spherically symmetric vacuum, up to second-order derivatives and quadratic in first-order derivative terms, is given by \cite{EMGcov}
\begin{widetext}
\begin{eqnarray}
    \tilde{H}_{\rm grav}
    &=& - \sqrt{E^x} \frac{\chi}{2} \bigg[ E^\varphi \bigg( \frac{\alpha_0}{E^x}
    + 2 \frac{\sin^2 \left(\lambda K_\varphi\right)}{\lambda^2}\frac{\partial c_{f}}{\partial E^x}
    + 4 \frac{\sin \left(2 \lambda K_\varphi\right)}{2 \lambda} \frac{1}{\lambda} \frac{\partial \left(\lambda q\right)}{\partial E^x}
    \nonumber\\
    &&\qquad\qquad
    + \left(\frac{\alpha_2}{E^x} - 2 \frac{\partial \ln \lambda^2}{\partial E^x}\right) \left( c_f \frac{\sin^2 \left(\lambda K_\varphi\right)}{\lambda^2}
    + 2 q \frac{\sin \left(2 \lambda K_\varphi\right)}{2 \lambda} \right)
    \nonumber\\
    &&\qquad\qquad
    + 4 \left(\frac{K_x}{E^\varphi} + \frac{K_\varphi}{2} \frac{\partial \ln \lambda^2}{\partial E^x} \right) \left(c_f \frac{\sin (2 \lambda K_\varphi)}{2 \lambda}
    + q \cos(2 \lambda K_\varphi)\right)
    \bigg)
    \nonumber\\
    &&\qquad\qquad
    + \frac{((E^x)')^2}{E^\varphi} \bigg(
    - \frac{\alpha_2}{4 E^x} \cos^2 \left( \lambda K_\varphi \right)
    + \left( \frac{K_x}{E^\varphi} + \frac{K_\varphi}{2} \frac{\partial \ln \lambda^2}{\partial E^x} \right) \lambda^2 \frac{\sin \left(2 \lambda K_\varphi \right)}{2 \lambda} \bigg)
    \nonumber\\
    &&\qquad\qquad
    + \left(\frac{(E^x)' (E^\varphi)'}{(E^\varphi)^2}
    - \frac{(E^x)''}{E^\varphi}\right) \cos^2 \left( \lambda K_\varphi \right)
    \bigg]
    \ ,
    \label{eq:Hamiltonian constraint - modified - non-periodic}
\end{eqnarray}
with the associated structure function 
\begin{eqnarray}
    \tilde{q}^{x x} &=&
    \left(
    \left( c_f
    + \lambda^2 \left( \frac{(E^x)'}{2 E^\varphi} \right)^2
    \right)
    \cos^2 \left( \lambda K_\varphi \right)
    - 2 q \lambda^2 \frac{\sin (2 \lambda K_\varphi)}{2 \lambda}
    \right) \chi^2
    \frac{E^x}{(E^\varphi)^2}
    =: \beta \frac{E^x}{(E^\varphi)^2}
    \,.
    \label{eq:Structure function - modified - non-periodic}
\end{eqnarray}
\end{widetext}
Here, $\chi , c_f , \alpha_0, \alpha_2, \lambda$, and $q$ are undetermined functions of $E^x$.
The classical constraint is recovered in the limit $\chi , c_{f} , \alpha_0, \alpha_2 \to 1$ and $\lambda, q \to 0$. (The cosmological constant can be recovered by instead setting $\alpha_{0} \to 1 - \Lambda E^x$, with $\Lambda>0$ corresponding to asymptotically de Sitter space in the classical limit.)

This vacuum system has the mass observable \cite{EMGscalar}
\begin{widetext}
\begin{eqnarray}
    \mathcal{M}
    &=&
    d_0
    + \frac{d_2}{2} \left(\exp \int {\rm d} E^x \ \left(\frac{\alpha_2}{2 E^x}
    - \frac{\partial \ln \lambda^2}{\partial E^x}\right)\right)
    \left(
    c_f \frac{\sin^2\left(\lambda K_{\varphi}\right)}{\lambda^2}
    + 2 q \frac{\sin \left(2 \lambda  K_{\varphi}\right)}{2 \lambda}
    - \cos^2 (\lambda K_\varphi) \left(\frac{\lambda (E^x)'}{2 E^\varphi}\right)^2
    \right)
    \nonumber\\
    &&
    + \frac{d_2}{4} \int {\rm d} E^x \ \left(\frac{\alpha_0}{E^x} \exp \int {\rm d} E^x \ \left(\frac{\alpha_2}{2 E^x} - \frac{\partial \ln \lambda^2}{\partial E^x}\right)\right)
    \,,
    \label{eq:Gravitational weak observable - DF - roots}
\end{eqnarray}
\end{widetext}
where $d_0$ and $d_2$ are constants. This expression is invariant under gauge transformations generated by the diffeomorphism and Hamiltonian constraints.

\subsubsection{Near minimum radius surface}
\label{app:Near minimum radius surface}

The minimum-radius coordinate $x=x_\lambda$ is defined as the largest solution to
\begin{equation}\label{eq:Minimum radius def}
    \beta (x_\lambda) = 0\,,
\end{equation}
or
\begin{equation}
1 + \lambda(x_\lambda)^2 \left( 1 - \frac{2 M}{x_\lambda} \right)=0\,.
\end{equation}
It follows that 
\begin{equation} \label{xlambda}
    x_{\lambda}=2M \frac{\lambda(x_{\lambda})^2}{1+\lambda(x_{\lambda})^2}
\end{equation}
and therefore $x_\lambda<x_{\rm H}$ for any $\lambda(x)$. (For non-constant $\lambda$, (\ref{xlambda}) is an implicit equation for $x_{\lambda}$. In particular, $x_{\lambda}$ is determined by a cubic polynomial equation for $\lambda^2=\Delta/x^2$.)

While multiple solutions may exist to Eq.~(\ref{eq:Minimum radius def}), only the largest one is relevant for the following discussions.
In an extension of spacetime across the coordinate singularity, the surface $x=x_\lambda$ is one of reflection symmetry beyond which $x$ starts increasing once again such that no radius lower than $x_\lambda$ exists: This shows that $x_\lambda$ is the minimum radius of the spacetime, which is always hidden beyond the horizon.
Furthermore, a special gauge exists \cite{ELBH} that can cross the surface $x=x_\lambda$ and in which the metric remains regular, as well as the Ricci and Kretschmann scalars, indicating that the geometry is indeed nonsingular.
The resulting global spacetime has the structure of an interuniversal wormhole, see Fig.~\ref{fig:Holonomy_KS_Vacuum_Wormhole-Periodic} for the conformal diagram of such a spacetime~\cite{ELBH}.

Expanding around this minimum radius surface as $x=x_\lambda+r$ to leading order in $r/x_\lambda$, we obtain the metric components
\begin{equation}
    1-\frac{2M}{x} \approx
    1-\frac{2M}{x_\lambda}
    + \frac{2M}{x_\lambda} \frac{r}{x_\lambda}
    \end{equation}
    as well as
    \begin{eqnarray}
    \beta &\approx& \chi^2 \left[ \lambda_\lambda^2 \frac{2M}{x_\lambda}
    + x_\lambda \left(1-\frac{2M}{x_\lambda}\right) \frac{\partial \lambda^2}{\partial x}\bigg|_{x_\lambda}
    \right] \frac{r}{x_\lambda}
    \nonumber\\
    &=:& \gamma \frac{r}{x_\lambda}\,,
    \end{eqnarray}
where $\gamma$ is positive since $\beta(x)>0$ for $x>x_\lambda$, and
    \begin{equation}
    \bar{\tilde{q}}^{xx} \approx \gamma \left(1-\frac{2M}{x_\lambda}\right) \frac{r}{x_\lambda}
    = - \frac{\gamma}{\lambda_\lambda^2} \frac{r}{x_\lambda}\,.
    \end{equation}
    Therefore,
    \begin{equation}
    \sqrt{\bar{N}^2 \bar{\tilde{q}}^{xx}}
    = \left(1-\frac{2M}{x}\right) \beta (x) \approx - \sqrt{\frac{\gamma}{x_\lambda}} \frac{r^{1/2}}{\lambda_\lambda^2}
    \end{equation}
    and
    \begin{equation}
    {\rm d} x^* \approx - \lambda_\lambda^2 \sqrt{\frac{x_\lambda}{\gamma}}\; r^{-1/2} {\rm d} r
    \end{equation}
    which integrates to
\begin{equation}
     \label{eq:Tortoise minimum radius}
   x^* \approx - 2 \lambda_\lambda^2 \sqrt{\frac{x_\lambda}{\gamma}}\; r^{1/2}
    + c_*
\end{equation}
where $\lambda_\lambda=\lambda(x_\lambda)$ and $c_*$ is the integration constant.
The metric near the minimum radius surface then takes the form
\begin{eqnarray}
    {\rm d} s^2 &\approx&
    \left(\frac{2M}{x_\lambda} - 1
    - \frac{2M}{x_\lambda} \frac{r}{x_\lambda}\right) {\rm d} t^2
    - \frac{\lambda_\lambda^2}{\gamma} \frac{x_\lambda}{r} {\rm d} r^2
    \nonumber\\
    &&+ x_\lambda^2 \left(1 + \frac{2 r}{x_\lambda}\right) {\rm d} \Omega^2
    \nonumber\\
    &\approx&
    \frac{2M}{x_\lambda} {\rm d} t^2
    - \frac{\lambda_\lambda^2}{\gamma} \frac{x_\lambda}{r} {\rm d} r^2
    + x_\lambda^2 {\rm d} \Omega^2
    \,,
    \label{eq:Spacetime metric - modified - Schwarzschild - near minimum radius}
\end{eqnarray}
where we have assumed $2M/x_\lambda\gg1$ in the second approximation.
Note that the classical limit of the expanded (\ref{eq:Spacetime metric - modified - Schwarzschild - near minimum radius}) cannot be defined because it is necessary to assume finite $x_\lambda$ to expand in $r/x_\lambda$.

For the constant case $\lambda(x)=\bar{\lambda}$ with $\chi=1/\sqrt{1+\bar{\lambda}^2}$ we obtain
\begin{eqnarray}
    \gamma_{\bar{\lambda}} \!\!&=&\!\!
    1
    \,,\\
    \lambda_{\bar{\lambda}}^2 \!\!&=&\!\! \frac{x_{\bar{\lambda}}}{2M-x_{\bar{\lambda}}}
    \,.
\end{eqnarray}
and for the decreasing case $\lambda(x)=\sqrt{\Delta/x^2}$ with $\chi=1$ we obtain
\begin{eqnarray}
    \gamma_\Delta \!\!&=&\!\!
    2 \frac{3M-x_\Delta}{2M-x_\Delta}
    \,,\\
    \lambda_\Delta^2 \!\!&=&\!\!
    \frac{x_\Delta}{2M-x_\Delta}
    \,.
\end{eqnarray}

The result (\ref{eq:Spacetime metric - modified - Schwarzschild - near minimum radius}) differs significantly from the classical metric near the singularity:
\begin{eqnarray}
    {\rm d} s^2 \approx
    \frac{2M}{r} {\rm d} t^2
    - \frac{r}{2M} {\rm d} r^2
    + r^2 {\rm d} \Omega^2
    \,.
    \label{eq:Spacetime metric - classical - Schwarzschild - near minimum radius}
\end{eqnarray}
Also, note that classically we have $x^*=x+2M\ln(x/(2M)-1)$ which cannot be extended to $x<2M$ with real values. (The full real range of  $-\infty<x^*<+\infty$ covers only the classical exterior outside of the horizon.) The interior can be reached only by complex continuation, which would result in a complex expansion $x^*\approx - r^2/(4M)+2i\pi M$ of the classical tortoise coordinate in terms of  $x=r$ near the singularity.
Compared with (\ref{eq:Tortoise minimum radius}), this expansion has the inverse exponent of $r$.

\subsection{Scalar matter nonminimal coupling}
\label{app:Scalar matter coupling}

\subsubsection{Canonical system}
\label{app:Scalar canonical system}

A general class of covariant, nonminimally coupled scalar matter was formulated in \cite{EMGscalar} for spherical symmetry.
The general constraint is given by
\begin{widetext}
\begin{eqnarray}
    \tilde{H} &=&
    -\chi \frac{\sqrt{E^x}}{2} \Bigg[
    E^\varphi \left(
    \frac{\alpha_0}{E^x} + \frac{\sin^2 (\lambda K_\varphi)}{\lambda^2} \left( \left(\frac{\alpha_2}{E^x} - 4 \frac{\partial \ln \lambda}{\partial E^x}\right) c_f + 2 \frac{\partial c_f}{\partial E^x} \right)\right)
    \nonumber\\
    &&\qquad\qquad
    + 2 E^\varphi \frac{\sin (2 \lambda K_\varphi)}{2 \lambda} \left( \left(\frac{\alpha_2}{E^x} - 2 \frac{\partial \ln \lambda}{\partial E^x}\right) q
    + 2 \frac{\partial q}{\partial E^x} \right)
    \nonumber\\
    &&\qquad\qquad
    + 4 E^\varphi \left( \frac{K_x}{E^\varphi}
    + \frac{\partial \ln \lambda}{\partial E^x} K_\varphi
    + \frac{P_\phi}{E^\varphi} c_{h3} \right)
    \left( \left(c_f+\lambda^2 \left(\frac{(E^x)'}{2E^\varphi}\right)^2\right) \frac{\sin (2 \lambda K_\varphi)}{2 \lambda}
    + q \cos(2 \lambda K_\varphi) \right)
    \nonumber\\
    &&\qquad\qquad
    - \frac{P_\phi{}^2}{E^\varphi} \frac{\alpha_3}{E^x} \left( \left(c_f+\lambda^2 \left(\frac{(E^x)'}{2E^\varphi}\right)^2\right) \cos^2 (\lambda K_\varphi)
    - 2 q \lambda^2 \frac{\sin (2 \lambda K_\varphi)}{2 \lambda} \right) \nonumber\\
    &&\qquad\qquad
    - \frac{((E^x)')^2}{E^\varphi} \frac{\alpha_2}{4 E^x} \cos^2 (\lambda K_{\varphi})
    + \left( \frac{(E^x)' (E^\varphi)'}{(E^\varphi)^2}
    - \frac{(E^x)''}{E^\varphi} \right) \cos^2 (\lambda K_\varphi)
    - \frac{1}{E^\varphi} \left( \phi' + c_{h3} (E^x)' \right)^2 \frac{E^x}{\alpha_3}
    - 2 E^\varphi V
    \Bigg]
    \nonumber\\
    &&
    + \chi^2 E^x \sqrt{\tilde{q}_{xx}} V_q
    + (E^\varphi)^2 \sqrt{\tilde{q}^{xx}} V^q
    \ ,
    \label{eq:Hamiltonian constraint - DF - scalar polymerization - roots}
\end{eqnarray}
with structure function 
\begin{eqnarray}
    \tilde{q}^{x x}
    =
    \chi^2 \left( \left( c_f
    + \left(\frac{\lambda (E^x)'}{2 E^\varphi}\right)^2 \right) \cos^2\left(\lambda K_{\varphi}\right)
    - 2 \lambda^2 q \frac{\sin \left(2 \lambda K_{\varphi}\right)}{2 \lambda} \right) \frac{E^x}{(E^\varphi)^2}
    \,.
    \label{eq:Structure function - DF - roots}
\end{eqnarray}
\end{widetext}
The new extra factor $c_{h3}$ is an undetermined function of $E^x$, while $V$, $V_q$, and $V^q$ are undetermined functions of $E^x$ and $\phi$.
Additionally, $\chi$ is allowed to depend on $\phi$ too.
The classical limit of these new modification functions is given by $c_{h3},V_q,V^q,\to0$, while $V(E^x,\phi)\to V(\phi)$ takes the role of the classical scalar potential.

When $V=V_q=V^q=0$, the system has the scalar-field observable
\begin{eqnarray}
    G [\alpha] &=& \int {\rm d}^3 x\ \alpha P_\phi
    \ ,
    \label{eq:Scalar field symmetry generator - DF - roots}
\end{eqnarray}
where $\alpha$ is a constant.
The associated conserved matter current $J^\mu$ has the components
\begin{eqnarray}
    J^t &=& P_\phi
    \ , \\
    J^x
    &=& \chi \frac{(E^x)^{3/2}}{\alpha_3 E^\varphi} \left( \phi'
    + c_{h3} (E^x)' \right)
    \ .
    \label{eq:Conserved matter current - DF - roots}
\end{eqnarray}

\subsubsection{Scalar matter coupling and perturbative dynamics: No backreaction}
\label{sec:Perturbation theory}

Consider a general scalar matter coupling in spherical symmetry with a Hamiltonian constraint of the form (\ref{eq:Ham const matter general}).
Consider now a perturbative scheme where we replace
\begin{eqnarray}
    E^x= E^x_{(0)}+E^x_{(1)} &,& K_x= K_x^{(0)}+K_x^{(1)}\,,\\
    E^\varphi= E^\varphi_{(0)}+E^\varphi_{(1)} &,& K_\varphi= K_\varphi^{(0)}+K_\varphi^{(1)}\,,\\
    \phi= \phi^{(0)}+\phi^{(1)} &,& P_\phi= P_\phi^{(0)}+P_\phi^{(1)}\,,
\end{eqnarray}
where the quantities indexed by $(0)$ correspond to the background variables (usually given by some simple solutions known in analytic form), while those indexed by $(1)$ are the perturbations. Relevant effects of perturbations in general appear through their first-order contributions to equations of motion, which are then linear in all perturbation variables. In canonical form, equations of motion are derived from Poisson brackets with the constraints. Since Poisson brackets include derivatives of phase-space functions by perturbation variables, the constraints should be expanded to quadratic order in perturbations such that all linear terms are included in equations of motion.
In addition, the constraints depend on the lapse function and shift vector, which are not directly subject to equations of motion but are often determined by consistency or gauge conditions. Therefore, we also need
 perturbation of the lapse and shift,
\begin{equation}
    N = N^{(0)} + N^{(1)} \quad,\quad
    N^x = N^x_{(0)} + N^x_{(1)}
    \,.
\end{equation}

The perturbed constraints then receive multiple contributions of the form
\begin{eqnarray}
    H_x[N^x] &=& H_x^{(0)}[N^x_{(0)}] + H_x^{(2)}[N^x_{(0)}] + H_x^{(1)}[N^x_{(1)}]
    \nonumber\\
    &=:& \bar{H}_x[N^x_{(0)}] + H_x^{(1)}[N^x_{(1)}]
    \end{eqnarray}
    for the diffeomorphism constraint, where
    \begin{eqnarray}
    H_x^{(0)} \!\!&=&\!\! E^\varphi_{(0)} (K_\varphi^{(0)})' - K_x^{(0)} (E^x_{(0)})' + P_\phi^{(0)} (\phi^{(0)})'
    \,\\
    H_x^{(2)} \!\!&=&\!\! E^\varphi_{(1)} (K_\varphi^{(1)})' - K_x^{(1)} (E^x_{(1)})' + P_\phi^{(1)} (\phi^{(1)})'
    \,\\
    H_x^{(1)} \!\!&=&\!\! E^\varphi_{(1)} (K_\varphi^{(0)})' - K_x^{(1)} (E^x_{(0)})' + P_\phi^{(1)} (\phi^{(0)})'
    \nonumber\\
    \!\!&&\!\!
    + E^\varphi_{(0)} (K_\varphi^{(1)})' - K_x^{(0)} (E^x_{(1)})' + P_\phi^{(0)} (\phi^{(1)})'\quad
\end{eqnarray}
and similarly for the contributions
\begin{eqnarray}
    \tilde{H}_{\rm grav}[N] &=& \tilde{H}_{\rm grav}^{(0)}[N^{(0)}] + \tilde{H}_{\rm grav}^{(2)}[N^{(0)}] + \tilde{H}_{\rm grav}^{(1)}[N^{(1)}]
    \nonumber\\
    &=& \bar{\tilde{H}}_{\rm grav}[N^{(0)}] +\tilde{H}_{\rm grav}^{(1)}[N^{(1)}]
\end{eqnarray}
and
\begin{eqnarray}
    \tilde{H}_{\rm scalar}[N] &=& \tilde{H}_{\rm scalar}^{(0)}[N^{(0)}] + \tilde{H}_{\rm scalar}^{(2)}[N^{(0)}] + \tilde{H}_{\rm scalar}^{(1)}[N^{(1)}]
    \nonumber\\
    &=& \bar{\tilde{H}}_{\rm scalar}[N^{(0)}] +\tilde{H}_{\rm scalar}^{(1)}[N^{(1)}]
\end{eqnarray}
to the Hamiltonian constraint.

Using these constraints also requires a perturbative expansion of the canonical structure, given by Poisson brackets derived from terms such as
\begin{equation}
    \int{\rm d}x (\dot{K}_x E^x+\dot{K}_{\varphi} E^{\varphi}+\dot{\phi}P_{\phi}) 
\end{equation}
in the canonical form of an action. Inserting the perturbed expressions for all variables does not necessarily result in a consistent canonical structure because mixed integrals such as $\int{\rm d}x \dot{K}_x^{(0)}E^x_{(1)}$ vanish only for special cases of background solutions. One example is given by perturbations that weaken symmetries of the background, such as a spatially homogeneous background with perturbative inhomogeneity. If the constant mode is removed from perturbations, mixed integrals are guaranteed to vanish, such that background and perturbations have independent canonical structures. However, in the present case, we do not assume different symmetry types for background and perturbation modes, which are both given by $x$-dependent functions if we use an expansion by spherical harmonics. The perturbations do not provide completely independent phase-space degrees of freedom for a generic background.

If we can find a perturbative setting such that background and perturbations are canonically independent, the canonical formalism is guaranteed to be consistent. In particular, since the perturbed system is a simple Taylor expansion of the original one, it remains anomaly-free and covariant to the given order, and no additional machinery, such as a derivation of restrictions on the non-perturbative modification functions, is needed to ensure consistency.
The only subtlety in the assumption that there is no backreaction of perturbations on the background lies in the dynamics, as it implies that the background and first-order constraint contributions, $\tilde{H}^{(0)}$ and $\tilde{H}^{(1)}$, must vanish independently on physical solutions. In contrast, the second-order contribution $\tilde{H}^{(2)}$ is neglected as a constraint contribution to the leading order. (It is still used to generate first-order equations of motion but not imposed as a contribution to constraint equations.)

Choosing a well-defined canonically independent background and perturbations is straightforward, even though they both have the same symmetry types. Since we consider the scalar field as a perturbation of the vacuum model and ignore backreaction, there are distinct physical degrees of freedom, given by the gravitational variables and matter, respectively.
The perturbed matter equations of motion will be generated only by the part of $\tilde{H}_{\rm scalar}^{(2)}[N^{(0)}]$ and $\tilde{H}_{\rm scalar}^{(2)}[N^x_{(0)}]$, which depends quadratically on the perturbed matter variables since
\begin{eqnarray}\label{eq:Matter grav pert decoupling}
    &&\left(\phi\right)^2 = \left(\phi^{(1)}\right)^2
    \quad,\quad \left(P_\phi\right)^2 = \left(P_\phi^{(1)}\right)^2
    \quad,\nonumber\\
    &&\phi P_\phi = \phi^{(1)} P_\phi^{(1)}
\end{eqnarray}
while higher-order contributions in a Taylor expansion vanish whenever $\phi^{(0)}=P_\phi^{(0)}=0$. Coefficients of these terms depend only on the background gravitational variables.
This observation, in turn, implies that $\tilde{H}^{(1)}=H_x^{(1)}=0$ and, therefore, the first order constraints are dynamically trivial and do not constrain the dynamics in the special case of a vacuum background.
Furthermore, the perturbed gravitational variables decouple from the perturbed matter variables in the equations of motion because of the expansion (\ref{eq:Matter grav pert decoupling}), making the former redundant because they result in trivial additions to the gravitational background variables and hence can be absorbed into the latter without loss of generality.
This procedure facilitates our following analysis because we only need consider the perturbations of the scalar matter and not the gravitational ones.
Formally, we can therefore set $K_x^{(1)}=K_{\varphi}^{(1)}=E^x_{(1)}=E^{\varphi}_{(1)}=0$ because they decouple from the matter perturbations, as well as  $\phi^{(0)}=P_\phi^{(0)}=0$ because we have a vacuum background. It is then clear that the non-zero background and perturbation  variables are canonically independent, and we have a well-defined canonical structure given by the Poisson brackets
\begin{eqnarray}
    \{K_x^{(0)}(x),E^x_{(0)}(y)\} = \{K_\varphi^{(0)}(x),E^\varphi_{(0)}(y)\}&&
    \nonumber\\
    = \{\phi^{(1)}(x),P_\phi^{(1)}(y)\} = \delta(x-y)&&
\end{eqnarray}

Specifically, the background solution for the modified systems is given in Subsection~\ref{Sec:Schwarzschild coordinates}, which, by the assumption of negligible backreaction, will not be affected by the perturbative dynamics.
Since we have fixed $\phi^{(0)}=P_\phi^{(0)}=0$ for the background, in the rest of this work we drop the index $(1)$ of the perturbed matter variables as it is understood that they are perturbations on a vacuum background.
Similarly, we drop the superindex $(0)$ in the gravitational variables and set them equal to the barred functions defined in Subsection~\ref{sec:Vacuum background}.

\section{Spectra of quasinormal modes}

\subsection{Standard GR}
\label{Sec:Standard case}

Note that the potential \eqref{eq:Minimal pot} in the 1-D effective Schr\"odinger equation \eqref{eq:KG minimal - EMG01} vanishes at the horizon. Thus, the Schr\"odinger-like equation reduces to a harmonic oscillator problem, whose solutions are:\\
$$
\tilde{u}_{lm}(x^*)=c_{1} e^{-i \omega x^{*}}+c_{2} e^{+i \omega x^{*}}, \quad x \rightarrow x_{\rm H}
$$
where $x_{\rm H}=2 M$.
The first of term is interpreted as an ingoing wave, i.e., a wave that travels inward and eventually falls into the black hole event horizon. The second term is interpreted as an outgoing wave, i.e., a wave that travels outward with respect to the black hole and can escape to spatial infinity. As these are travelling waves, the second term would represent waves coming from the interior of the black hole. Since the perturbations are strictly classical, nothing is expected to come out from the black hole interior, thus, in the following analysis we impose the first term as the boundary condition at the horizon, which is accomplished by setting $c_{2}=0$.

We can do a similar investigation near spatial infinity, at $x \rightarrow \infty$, where $\bar{N}(x) \rightarrow 1$ and the effective potential  \eqref{eq:Minimal pot} vanishes. Thus, in such a limit the general solution to the wave equation \eqref{eq:KG minimal - EMG01} has the same form as the function given above:
$$
\tilde{u}_{lm}(x^*) = c_{3} e^{-i \omega x^{*}}+c_{4} e^{i \omega x^{*}}, \quad x \rightarrow \infty . $$
The first term is interpreted physically as waves coming in from outside the universe and must be avoided by setting $c_{3}=0$. In turn, the second term represents waves going out of the universe, this is the boundary condition at the spatial infinity. Finally, note that the boundary conditions do not depend explicitly on the angular momentum $\ell$. 

Using the relation between the radial coordinate and tortoise coordinate for a background spacetime~\eqref{eq:Schwarzschild}, we have:
$$ 
x^{*}=\int \frac{{\rm d} x}{f^{\prime}\left(x_{\rm H}\right)\left(x-x_{\rm H}\right)} \approx \frac{\ln \left(x-x_{\rm H}\right)}{f^{\prime}\left(x_{\rm H}\right)}, \quad x \rightarrow x_{\rm H} $$
Thus, in terms of the radial coordinate, the boundary condition at the horizon becomes $\left(f^{\prime}\left(x_{\rm H}\right)=1 / x_{\rm H}\right)$
\begin{equation}\label{eq:Boundary cond - horizon - standard}
\tilde{u}_{lm}(x) \sim e^{-i \omega \frac{\ln \left(x-x_{\rm H}\right)}{f^{\prime}\left(x_{\rm H}\right)}} \sim\left(x-x_{\rm H}\right)^{-i \omega x_{\rm H}} \,.
\end{equation}
Similarly, we can write the boundary condition at the spatial infinity, using the following relation:
$$ 
x^{*}=\int \frac{{\rm d} x}{f(x)} \approx x+x_{\rm H} \ln x, \quad x \rightarrow \infty \,.
$$
Hence, the asymptotic solution at the spatial infinity becomes:
\begin{equation}\label{eq:Boundary cond - Inf - standard}
    \tilde{u}_{lm}(x) \sim e^{i \omega\left(x + x_{\rm H} \ln x \right)} \sim x^{i \omega x_{\rm H}} e^{+i \omega x} \, . 
\end{equation}

The eigenvalue problem in equation (\ref{eq:KG minimal - EMG01}) is formulated to solve for the square of the frequency, denoted as $\omega^2$. This is consistent with the WKB procedure used in the following subsection for EMG. To obtain the actual frequencies, we need to take the square root, resulting in two branches: $\pm \sqrt{\omega^2}$. In the case of complex frequencies $\omega=\omega_R + i \omega_I$, we define the positive branch as the one with ${\rm sgn}(\omega_R)=+1$, denoted as $\omega^+$. Similarly, the negative branch is given by $\omega^-=-\omega^+=-{\rm Re}[\omega^+]-i {\rm Im}[\omega^+]$.

When we substitute the eigenfrequency into the asymptotic value (\ref{eq:Boundary cond - Inf - standard}), we obtain $\tilde{u}_{lm}^\pm(x)\sim e^{+i\omega^\pm x}\sim e^{-{\rm Im}[\omega^\pm]x}$, where the $\pm$ subindex denotes the respective branch solution. It is evident that the mode will be negligible if ${\rm sgn}\left({\rm Im}[\omega^\pm]\right)=+1$, and will be observable only if ${\rm sgn}\left({\rm Im}[\omega^\pm]\right)=-1$.
For instance, in GR, the fundamental scalar mode has the eigenfrequencies $x_{\rm H} \omega_{00}^+\approx0.209294-i0.230394$ and $x_{\rm H} \omega_{00}^-\approx-0.209294+i0.230394$, as obtained from a WKB method to third order \cite{Moreira_2023}; in this case, only the positive branch mode with eigenfrequency $\omega_{00}^+$ is observable in the asymptotic region, while the mode with $\omega_{00}^-$ is negligible.

As we will show below, in more complicated gravitational theories, including EMG, it is possible to obtain an eigenfrequency whose real and imaginary parts are both negative; in such a case, the mode is not negligible in the asymptotic region and hence observable.
Notice, in particular, that the mode near the horizon, $\tilde{u}_{lm}^\pm\sim e^{- i \omega^\pm (t+x^*)}$, remains ingoing regardless of the chosen branch and the mode in the asymptotic region, $\tilde{u}_{lm}^\pm\sim e^{- i \omega^\pm (t-x^*)}$, remains outgoing regardless of signs of $\omega^\pm$. Therefore, both branches can satisfy the boundary conditions. The sign in the real frequency component need not raise concerns because it goes unnoticed in measurements (as opposed to a wave number, which determines outgoing versus incoming modes). To take into account the possibility of the negative branch modes contributing to observations in the asymptotic region, we define the observable frequency spectrum as $\{\omega_{nl}^{\rm (obs)}\}=\{\omega_{nl}^+ \cup \omega_{nl}^-| {\rm sgn}\left({\rm Im}[\omega_{nl}^\pm]\right) = -1 \}$, that is, the union of the two branches that survive in the asymptotic region.

\end{appendix}

\newcommand{\noopsort}[1]{}


\begin{thebibliography}{10}

\bibitem{GW1}
The LIGO Scientific Collaboration, and The Virgo Collaboration, Phys.\ Rev.\ Lett. {\bf 116}, 061102 (2016); B. P. Abbott et al. [LIGO Scientific and Virgo Collaborations],
Phys.\ Rev.\ Lett.\ {\bf 119}, 161101 (2017); 
 B.~P.~Abbott {\it et al.} [LIGO Scientific and Virgo Collaborations],
  Phys.\ Rev.\ Lett.\  {\bf 119}, no. 14, 141101 (2017)
B.~P.~Abbott {\it et al.} [LIGO Scientific and Virgo Collaborations],
  Phys.\ Rev.\ X {\bf 9}, no. 3, 031040 (2019)
  

\bibitem{GW2}
The LIGO Scientific Collaboration, and The Virgo Collaboration, Phys.\ Rev.\ Lett. {\bf 116}, 241103 (2016)

\bibitem{GW_Review}
S.~Chandrasekhar, {\it The Mathematical Theory of Black Holes}, Oxford University Press (1983); 
H.~P.~Nollert,
  Class.\ Quant.\ Grav.\  {\bf 16}, R159 (1999);
K.~D. Kokkotas and B.~G. Schmidt, Living\ Rev.\ Rel.\ 2, {\bf 2} (1999), arXiv:9909058.
R.~A.~Konoplya and A.~Zhidenko,
  Rev.\ Mod.\ Phys.\  {\bf 83}, 793 (2011)

\bibitem{ReggeWheeler}
T. Regge and J.~A. Wheeler, Phys.\ Rev.\ D {\bf 108}, 1063 (1957).

\bibitem{Zerilli1}
F.~J. Zerilli, Phys.\ Rev.\ Lett. {\bf 24}, 737 (1970).

\bibitem{Zerilli2}
F.~J. Zerilli, Phys.\ Rev.\ D {\bf 2}, 2141 (1970).

\bibitem{ChandraQNM}
S. Chandrasekhar and S. Detweiler, Proc.\ R.\ Soc.\ Lond.\ A {\bf 344}, 411-452 (1975)

\bibitem{Cardoso_2019}
V. Cardoso and P. Pani, Living\ Rev\ Relativ {\bf 22}, 4 (2019).

\bibitem{Baibhav_2023}
V. Baibhav, {\it et al.}, Phys.\ Rev.\ D {\bf 108}, 104020 (2023).

\bibitem{GW170104} 
  B.~P.~Abbott {\it et al.} [LIGO Scientific and VIRGO Collaborations],
  Phys.\ Rev.\ Lett.\  {\bf 118}, no. 22, 221101 (2017)


 \bibitem{TestofGR}
 E.~Berti et al., Class.\ Quant.\ Grav.\ {\bf 32}, 243001 (2015);
 Nicolas Yunes, Kent Yagi, and Frans Pretorius, Phys.\ Rev.\ {\bf D94}, 084002 (2016).

\bibitem{ToG}
B.~P. Abbott et al. (Virgo, LIGO Scientific Collaborations), Phys.\ Rev.\ Lett.\ {\bf 116}, 221101 (2016).  

\bibitem{GWTemplates}
B.~P. Abbott et al. (Virgo, LIGO Scientific), Phys.\ Rev.\ {\bf D93}, 122003 (2016).

\bibitem{NoHair}
V. Cardoso and L. Gualtieri, Class. Quant. Grav. {\bf 33}, 174001 (2016)
 
\bibitem{Evans:2016mbw}
B.~P.~Abbott {\it et al.} [LIGO Scientific Collaboration],
  Class.\ Quant.\ Grav.\  {\bf 34} (2017) no.4,  044001
  doi:10.1088/1361-6382/aa51f4, arXiv:1607.08697 [astro-ph.IM].

\bibitem{Evans:2023euw}
M.~Evans, A.~Corsi, C.~Afle, A.~Ananyeva, K.~G.~Arun, S.~Ballmer, A.~Bandopadhyay, L.~Barsotti, M.~Baryakhtar and E.~Berger, \textit{et al.}
[arXiv:2306.13745 [astro-ph.IM]].

\bibitem{Maggiore:2019uih}
M.~Maggiore, C.~Van Den Broeck, N.~Bartolo, E.~Belgacem, D.~Bertacca, M.~A.~Bizouard, M.~Branchesi, S.~Clesse, S.~Foffa and J.~Garc\'\i{}a-Bellido, \textit{et al.}
JCAP \textbf{03}, 050 (2020)
arXiv:1912.02622 [astro-ph.CO].

\bibitem{Branchesi:2023mws}
M.~Branchesi, M.~Maggiore, D.~Alonso, C.~Badger, B.~Banerjee, F.~Beirnaert, E.~Belgacem, S.~Bhagwat, G.~Boileau and S.~Borhanian, \textit{et al.}
JCAP \textbf{07}, 068 (2023)
[arXiv:2303.15923 [gr-qc]].

\bibitem{LISA}
P. Amaro-Seoane, {\it et al.} [LISA Collaboration], \textit{Laser interferometer space antenna}, (2017), arXiv:1702.00786.

\bibitem{Barausse:2020rsu}
E.~Barausse, E.~Berti, T.~Hertog, S.~A.~Hughes, P.~Jetzer, P.~Pani, T.~P.~Sotiriou, N.~Tamanini, H.~Witek and K.~Yagi, \textit{et al.}
Gen. Rel. Grav. \textbf{52}, no.8, 81 (2020)
doi:10.1007/s10714-020-02691-1
[arXiv:2001.09793 [gr-qc]].

\bibitem{LISA:2022kgy}
K.~G.~Arun \textit{et al.} [LISA],
Living Rev. Rel. \textbf{25}, no.1, 4 (2022)
doi:10.1007/s41114-022-00036-9
[arXiv:2205.01597 [gr-qc]].

\bibitem{QNMSNRbound}
H.~Nakano, T.~Tanaka, and T.~Nakamura, Phys.\ Rev.\ D\ {\bf 92}, 064003 (2015)

\bibitem{Shanki-Joseph:Review}
S.~Shankaranarayanan and J.~P.~Johnson,
Gen. Rel. Grav. \textbf{54}, no.5, 44 (2022),
arXiv:2204.06533 [gr-qc].
 
\bibitem{2018-Barack.etal-CQG} 
  L.~Barack {\it et al.},
  Class.\ Quant.\ Grav.\  {\bf 36}, no. 14, 143001 (2019)

\bibitem{Shanki_QNM_fR}
S. Bhattacharyya and S. Shankaranarayanan, Eur.\ Phys.\ Journal\ C {\bf 78} 737 (2018)

\bibitem{Shanki-QNM-dCS}
S.~Bhattacharyya and S.~Shankaranarayanan,
Phys. Rev. D \textbf{100}, no.2, 024022 (2019),
arXiv:1812.00187 [gr-qc].


\bibitem{Shanki-Essay:2019}
S.~Shankaranarayanan,
Int. J. Mod. Phys. D \textbf{28}, no.14, 1944020 (2019),
arXiv:1905.03943 [gr-qc].


\bibitem{EMG}
M. Bojowald and E.~I. Duque, Class.\ Quant.\ Grav. \textbf{42}, 095008 (2024), arXiv:2404.06375.

\bibitem{EMGcov}
M. Bojowald and E.~I. Duque, Phys.\ Rev.\ D {\bf 108}, 084066, (2023), arXiv:2310.06798.

\bibitem{alonso2022nonsingular}
A. Alonso-Bardaj\'{\i}, D. Brizuela and R. Vera, Phys.\ Rev.\ D {\bf 106}, 024035 (2022), arXiv:2205.02098.

\bibitem{Alonso_Bardaji_2022}
A. Alonso-Bardaj\'{\i}, D. Brizuela and R. Vera, Phys.\ Lett.\ B {\bf 829}, 137075 (2022), arXiv:2112.12110.

\bibitem{ELBH}
I.~H. Belfaqih, M. Bojowald, S. Brahma, and E.~I. Duque,
\textit{Black holes in effective loop quantum gravity: Covariant holonomy modifications}, arXiv:2407.12087.

\bibitem{EMGPF}
E.~I. Duque, Phys.\ Rev.\ D {\bf 109}, 044014, (2024), 	arXiv:2311.08616.

\bibitem{milgrom1983modification}
M. Milgrom, Astrophysical Journal {\bf 270}, 371 (1983).

\bibitem{banik2022galactic}
B. Indranil and Z. Hongsheng, Symmetry {\bf 14}, 1331 (2022), arXiv:2110.06936v9.

\bibitem{MONDEMG}
M. Bojowald and E.~I. Duque, Phys.\ Lett.\ B {\bf 847}, 138279 (2023), arXiv:2310.19894.

\bibitem{alonsobardaji2023Charged}
A. Alonso-Bardaj\'{\i}, D. Brizuela, and R. Vera, Phys.\ Rev.\ D {\bf 107}, 064067 (2023), arXiv:2302.10619.

\bibitem{alonso2021anomaly}
A. Alonso-Bardaj\'{\i} and D. Brizuela, Phys.\ Rev.\ D {\bf 104}, 084064 (2021), arXiv:2106.07595.

\bibitem{EMGscalar}
M. Bojowald and E.~I. Duque, Phys.\ Rev.\ D {\bf 109}, 084006 (2024) arxiv:2311.10693.

\bibitem{EmEM}
E.~I. Duque, \textit{Emergent electromagnetism}, 	arXiv:2407.14954.

\bibitem{MinCoup}
A. Alonso-Bardaj\'{\i}, and D. Brizuela, Phys.\ Rev.\ D {\bf 109}, 044065 (2024), arXiv:2310.12951.

\bibitem{Fu_2023}
G. Fu, D. Zhang, P. Liu, X.~M. Kuang, and J.~P. Wu \newblock Phys.\ Rev.\ D {\bf 109} 026010 (2024)

\bibitem{Moreira_2023}
Z.~S. Moreira, H.~C.~D. Lima Junior, L.~C.~B. Crispino, and C.~A.~R. Herdeiro \newblock Phys.\ Rev.\ D {\bf 107} 104016 (2023)

\bibitem{Gingrich_2024}
D.~M. Gingrich \newblock Phys.\ Rev.\ D {\bf 110} 084045 (2024)

\bibitem{bojowald2000symmetry}
M. Bojowald and H. Kastrup, Class.\ Quant.\ Grav. {\bf 17}, 3009 (2000), arXiv:hep-th/9907042.

\bibitem{bojowald2004spherically}
M. Bojowald, Class.\ Quant.\ Grav. {\bf 21}, 3733 (2004), arXiv:gr-qc/0407017.

\bibitem{Das_2005}
S. Das and S. Shankaranarayanan, Class.\ Quantum\ Grav. {\bf 22} L7-L21 (2005), arXiv:hep-th/0410209.

\bibitem{IyerWKB}
S. Iyer and C.~M. Will, Phys.\ Rev.\ D {\bf 35}, 3621 (1987)

\bibitem{NISThandbook}
F.~Olver, D.~Lozier, R.~Boisvert, and C.~Clark, {\sl The NIST Handbook of Mathematical Functions}, Cambridge University Press, New York, NY (2010)

\bibitem{Kummereqs}
A. Milton and I.~A. Stegun, eds. {\sl Handbook of mathematical functions with formulas, graphs, and mathematical tables}, Vol. 55, US Government printing office, (1968)

\bibitem{Nietzke}
L. Motl and A. Neitzke, Adv. Theor. Math. Phys. {\bf 7}, 307-330 (2003), arXiv:hep-th/0301173.

\bibitem{BHbomb}
W. Press and S. Teukolsky, Nature {\bf 238}, 211-212 (1972)

\bibitem{QNMModified}
F. Moulin, A.  Barrau and K. Martineau, Universe {\bf 5}, 202 (2019), arXiv:1908.06311.

\bibitem{PertLQG}
M. Bojowald, G.~M. Hossain, M. Kagan, and S. Shankaranarayanan, Phys.\ Rev.\ D  {\bf 78}, 78.063547 (2006), arXiv:0806.3929.

\bibitem{Johnson:2023skw}
J.~P.~Johnson, S.~Jana and S.~Shankaranarayanan,
Phys. Rev. D \textbf{109}, no.2, L021501 (2024), 
arXiv:2303.13271 [gr-qc].

\end{thebibliography}
\end{document}